\documentclass[5p,sort&compress]{elsarticle}

\journal{Journal of \LaTeX\ Templates}

\usepackage{booktabs}
\usepackage{paralist}
\usepackage{xcolor}
\usepackage{amsmath}
\usepackage{hyperref}
\usepackage{multirow}
\usepackage{tabu}
\usepackage{diagbox}
\usepackage{gensymb}
\usepackage{siunitx}
\usepackage{mathtools}

\usepackage[linesnumbered, ruled]{algorithm2e}
\SetAlFnt{\small}
\SetAlCapFnt{\small}
\SetAlCapNameFnt{\small}
\SetKwRepeat{Do}{do}{while}

\usepackage[numbers]{natbib}
\usepackage{makecell}
\newcolumntype{P}[1]{>{\centering\arraybackslash}p{#1}}

\usepackage{algorithmic}

\usepackage{caption}
\usepackage{subcaption}

\usepackage{tcolorbox}

\begin{document}

\begin{frontmatter}

\title{Environment Imitation: Data-Driven Environment Model Generation Using Imitation Learning for Efficient CPS Goal Verification}

\author[kaist]{Yong-Jun Shin}
\ead{yjshin@se.kaist.ac.kr}

\author[lux]{Donghwan Shin\corref{corresponding}}
\cortext[corresponding]{Corresponding author}
\ead{donghwan.shin@uni.lu}

\author[kaist]{Doo-Hwan Bae}
\ead{bae@se.kaist.ac.kr}

\address[kaist]{Korea Advanced Institute of Science and Technology}
\address[lux]{University of Luxembourg}

\begin{abstract}
Cyber-Physical Systems (CPS) continuously interact with their physical environments through software controllers that observe the environments and determine actions. Engineers can verify to what extent the CPS under analysis can achieve given goals by analyzing its Field Operational Test (FOT) logs. However, it is challenging to repeat many FOTs to obtain statistically significant results due to its cost and risk in practice. To address this challenge, simulation-based verification can be a good alternative for efficient CPS goal verification, but it requires an accurate virtual environment model that can replace the real environment that interacts with the CPS in a closed loop.
This paper proposes a novel data-driven approach that automatically generates the virtual environment model from a small amount of FOT logs. We formally define the environment model generation problem and solve it using Imitation Learning (IL) algorithms. In addition, we propose three specific use cases of our approach in the evolutionary CPS development. To validate our approach, we conduct a case study using a simplified autonomous vehicle with a lane-keeping system. The case study results show that our approach can generate accurate virtual environment models for CPS goal verification at a low cost through simulations.
\end{abstract}


\end{frontmatter}

\section{Introduction}\label{sec:intro}

Cyber-Physical Systems (CPS) utilize both physical and software components deeply intertwined to continuously collect, analyze, and control physical actuators at runtime~\cite{baheti2011cyber}. CPS has been increasingly studied for many applications, such as autonomous vehicles~\cite{an2020uncertainty, mullins2018adaptive}, robots~\cite{bozhinoski2019safety, ahmad2016software}, smart factories~\cite{shiue2018real, wang2021proactive}, and medical devices~\cite{zema2015developing, fu2011trustworthy}. 

One of the essential problems in CPS development is to verify to what extent the CPS under development can achieve its goals. To answer this, a developer could deploy a CPS (e.g., an autonomous vehicle) into its operational environment (e.g., a highway road) and verify the CPS's goal achievement (e.g., lane-keeping) using the logs collected from the Field Operational Tests (FOTs). However, conducting FOTs is expensive, time-consuming, and even dangerous, especially when hundreds of repeats are required to achieve a certain level of statistical significance in the verification results. An alternative is a simulation-based approach where the software controller of the CPS is simulated with a virtual environment model. Though it can reduce the cost and risk of the CPS goal verification compared to using FOTs, it requires a highly crafted virtual environment model based on deep domain knowledge. Furthermore, it may not be possible at all if a high-fidelity simulator for the problem domain does not exist. It prevents the simulation-based approach from being better used in practice.

To solve the difficulty of manually generating virtual environment models, we propose an automated data-driven environment model generation approach for CPS goal verification by recasting the problem of environment model generation as the problem of imitation learning. We call this novel approach \textit{ENVironment Imitation} (\textit{ENVI}). In machine learning, Imitation Learning (IL) has been widely studied to mimic complex human behaviors in a given task only from a limited amount of demonstrations~\cite{hussein2017imitation}. Our approach leverages IL to mimic how the real environment interacts with the CPS under analysis from a small set of log data collected from FOTs. Since the log data records how the CPS and the real environment interacted, our approach can generate an environment model that mimics a state transition mechanism of the real environment according to the CPS action as closely as possible to that recorded in the log data. The generated environment model is then used to simulate the CPS software controller as many times as needed to statistically analyze the CPS goal achievement. 

We evaluate the feasibility of our novel approach while comparing various imitation learning algorithms on a case study of a lane-keeping system of an autonomous robot vehicle. The evaluation results show that our approach can automatically generate an environment model that mimics the interaction mechanism between the lane-keeping system and physical environment, even using minimal amounts of FOT log data (e.g., less than 30 seconds execution log).

In summary, below are the contributions of this paper: 
\begin{enumerate}[1)]
    \item We shed light on the problem of environment model generation for CPS goal verification with a formal problem definition.
    \item We propose \textit{ENVI}, a novel data-driven approach for environment model generation utilizing IL.
    \item We assess the application of our approach through a case study with a real CPS and various IL algorithms.
\end{enumerate}

The remainder of this paper is organized as follows: Section~\ref{sec:motivatingEx} illustrates a motivating example. Section~\ref{sec:background} provides background on representative imitation learning algorithms considered in our experiments. Section \ref{sec:problemDef}~formalizes the problem of the data-driven environment model generation. Section~\ref{sec:approach} describes the steps of \textit{ENVI}. Section~\ref{sec:caseStudy} reports on the evaluation of \textit{ENVI}. Section~\ref{sec:discussion} discusses implications and open issues. Section~\ref{sec:relatedWork} introduces related work. Section \ref{sec:conclusion} concludes the paper.

\section{Motivating Example}\label{sec:motivatingEx}
We present a simple example of CPS goal verification to demonstrate a use case of our approach.

Consider a software engineer developing a lane-keeping system of an autonomous vehicle. The engineer aims to develop and test the vehicle's software controller (i.e., lane-keeping system) that continuously monitors the distance from the center of the lane and computes the steering angle that determines how much to turn to keep the distance as small as possible.

Once the software controller is developed, the engineer must ensure that the vehicle equipped with the controller continues to follow the center of the lane while driving. To do this, the engineer deploys the vehicle on a safe road and collect an FOT log, including the distance $d_t$ and the steering angle $a_t$ at time $t=1,\dots,T$ where $T$ is a pre-defined FOT duration. Based on the collected data, the engineer can quantitatively assess the quality of the lane-keeping system by calculating the sum of the distances the vehicle deviated from the center of the lane, i.e., $\Sigma_{t=1}^{T} |d_t|$. The quantitative assessment is used to verify precisely a goal of the system, i.e., whether $\Sigma_{t=1}^{T} |d_t| < \epsilon$ holds or not for a small threshold $\epsilon$. Notice that, due to the uncertainties in FOT, such as non-uniform friction between the tires and the ground, the same FOT must be repeated multiple times, and statistical analysis should be applied to the results.

It takes a lot of time and resources to repeat the FOTs enough to obtain statistically significant results. To address this issue, the engineer may decide to rely on simulations. However, using high-fidelity and physics-based simulators, such as Webots~\cite{michel2004cyberbotics} or Gazebo~\cite{Koenig-2004-394}, is very challenging, especially for software engineers who do not have enough expertise in physics. It is not easy to accurately design the physical components of the system (e.g., the size of wheels and the wheelbase) and the road in the simulator so that the simulation results are almost identical to the FOT results. 

Our approach, \textit{ENVI}, enables the CPS goal verification without using such a high-fidelity simulator. The engineer can simply provide \textit{ENVI} with the software controller (i.e., the lane-keeping system under analysis) and a small amount of FOT logs collected from the beginning, which is far less than the data required for statistically significant results using FOTs. Then \textit{ENVI} automatically generates a virtual environment model that imitates the behavior of the real environment of the lane-keeping system; specifically, the virtual environment model can simulate $d_{t+1}$ for given $d_t$ and $a_t$ for $t=2,\dots,T$ such that $\Sigma_{t=1}^{T} |d_t|$ calculated based on the virtual model is almost the same as the value calculated based on the FOTs. Therefore, by quickly re-running the simulation multiple times, the engineer can have statistically significant results about the quality of the software controller at little cost. Furthermore, if multiple software controller versions make different CPS behaviors, the virtual environment model generated by \textit{ENVI} can be reused to verify the CPS goal achievements of new controller versions that have never been tested in the real environment.

The challenge for \textit{ENVI} is automatically generating a virtual environment model that behaves as similar as possible to the real environment using a limited amount of data. To address this, we leverage \textit{imitation learning} detailed in Section~\ref{sec:background}.

\section{Background: Imitation Learning}\label{sec:background}
Imitation Learning (IL) is a learning method that allows an agent to mimic expert behaviors for a specific task by observing demonstrations of the expert~\cite{hussein2017imitation}. For example, an autonomous vehicle can learn to drive by observing how a human driver controls a vehicle. IL assumes that an expert decides an action depending on only the state that the expert encounters. Based on this assumption, an expert demonstration is a series of pairs of states and actions, and IL aims to extract the expert’s internal decision-making function (i.e., a policy function that maps states into actions) from the demonstration~\cite{hussein2017imitation}. We introduce two representative IL algorithms in the following subsections: Behavior Cloning (BC) and Generative Adversarial Imitation Learning (GAIL).

\subsection{Behavior Cloning}\label{sec:background:bc}
Behavior Cloning (BC) infers the policy function of the expert using supervised learning~\cite{schaal1997learning, argall2009survey}. Training data can be organized by pairing states and corresponding actions in the expert's demonstration. Then existing supervised learning algorithms can train the policy function that returns expert-like actions for given states.
Due to the simplicity of the BC algorithm, BC can create a good policy function that mimics the expert quickly if there are sufficiently much demonstration data. 
However, if the training data (i.e., expert demonstration) does not fully cover the input state space or is biased, the policy function may not mimic the expert behavior correctly~\cite{argall2009survey}.

\subsection{Generative Adversarial Imitation Learning}\label{sec:background:gail}
Generative Adversarial Imitation Learning (GAIL)~\cite{ho2016generative} utilizes the idea of Generative Adversarial Networks~\cite{goodfellow2014generative} to evolve the policy function using iterative competitions with a \textit{discriminator} that evaluates the policy function.
Therefore, both the policy function and the discriminator are trained in parallel.

The policy function gets states in the expert demonstration and produces simulated actions. The discriminator then gets the policy function's input (i.e., states) and output (i.e., simulated actions) and evaluates how the policy function behaves like the real expert, as shown in the demonstration. The more similar the simulation is to the expert demonstration, the more rewarded the policy function is by the discriminator. The policy function is trained to maximize the reward from the discriminator.

On the other hand, the discriminator is trained using both the demonstration data and the simulation trace of the policy function. The state and action pairs, which is the input and output of the policy function, in the demonstration data are labeled as \textit{real}, but the pairs in the simulation trace are labeled as \textit{fake}. A supervised learning algorithm trains the discriminator to quantitatively evaluate whether a state and action pair is real (returning a high reward) or fake (returning a low reward). 

After numerous learning iterations of the policy function and the discriminator, the policy function finally mimics the expert well to deceive the advanced discriminator. GAIL uses both the expert demonstration data and the simulation trace data of the policy function generated internally, so it works well even with small demonstration data~\cite{ho2016generative}. However, because of the internal simulation of the policy function, its learning speed is relatively slow~\cite{jena2020augmenting}.

\section{Problem Definition}\label{sec:problemDef}
This section introduces a mathematical framework for modeling how the CPS under analysis interacts with its environment to achieve its goals. Based on the formal framework, we then define the environment model generation problem for CPS goal verification.

\subsection{A Formal Framework for CPS Goal Verification}
A CPS achieves its goals by interacting with its physical environment. Specifically, starting from an initial state of the environment, the CPS software controller observes the state and decides an appropriate action to maximize the likelihood of achieving the goals. Then, taking action causes a change in the environment for the next step, which the CPS will observe again to decide an action for the next step. We assume the CPS and the environment interact in a closed loop without interference by a third factor. To formalize this process, we present a novel \textit{CPS-ENV interaction model} inspired by Markov Decision Process~\cite{sutton1998introduction} that models an agent's sequential decision-making process under observation over its environmental states.

Specifically, a \textit{CPS-ENV interaction model} is a tuple $M = (S, A, \pi, \delta, s_0)$, where $S$ is a set of observable states of the environment under consideration, $A$ is a set of possible CPS actions, $\pi:S \rightarrow A$ is a policy function that captures the software controller of the CPS, $\delta:S\times A\rightarrow S$ is a transition function that captures the transitions of environmental states over time as a result of CPS actions and its previous states\footnote{Though we use deterministic policy and transition functions for simplicity, they can be easily extended in terms of probability density, i.e., $\pi: S\times A \rightarrow [0,1]$ and $\delta:S\times A\times S\rightarrow [0, 1]$, to represent stochastic behaviors if needed.\label{foot:formalism1}}, and $s_0$ is an initial environmental state.
For example, starting from $s_0$, the CPS makes an action $a_0 = \pi(s_0)$, leading to a next state $s_1 = \delta(s_0, a_0)$. By observing $s_1$, the CPS again makes the next action $a_1=\pi(s_1)$, and so on. 

For a \textit{CPS-ENV interaction model} $M = (S, A, \pi, \delta, s_0)$, we can think of a sequence of transitions $s_0\xrightarrow{a_0}s_1\xrightarrow{a_1}s_2\xrightarrow{a_2}...\xrightarrow{a_{n-1}}s_n$ over $n$ steps where $s_{t-1}\xrightarrow{a_{t-1}}s_t$ denotes a transition from a state $s_{t-1}$ to another state $s_t$ of the environment by taking an action $a_{t-1}$ of the CPS. More formally, we define a \textit{trajectory} of $M$ over $T$ time ticks as a sequence of tuples $\mathit{tr}(M, T) = \langle (s_0, a_0), \dots, (s_
T, a_T) \rangle$. 

\begin{figure}
    \centering
    \includegraphics[width=\linewidth]{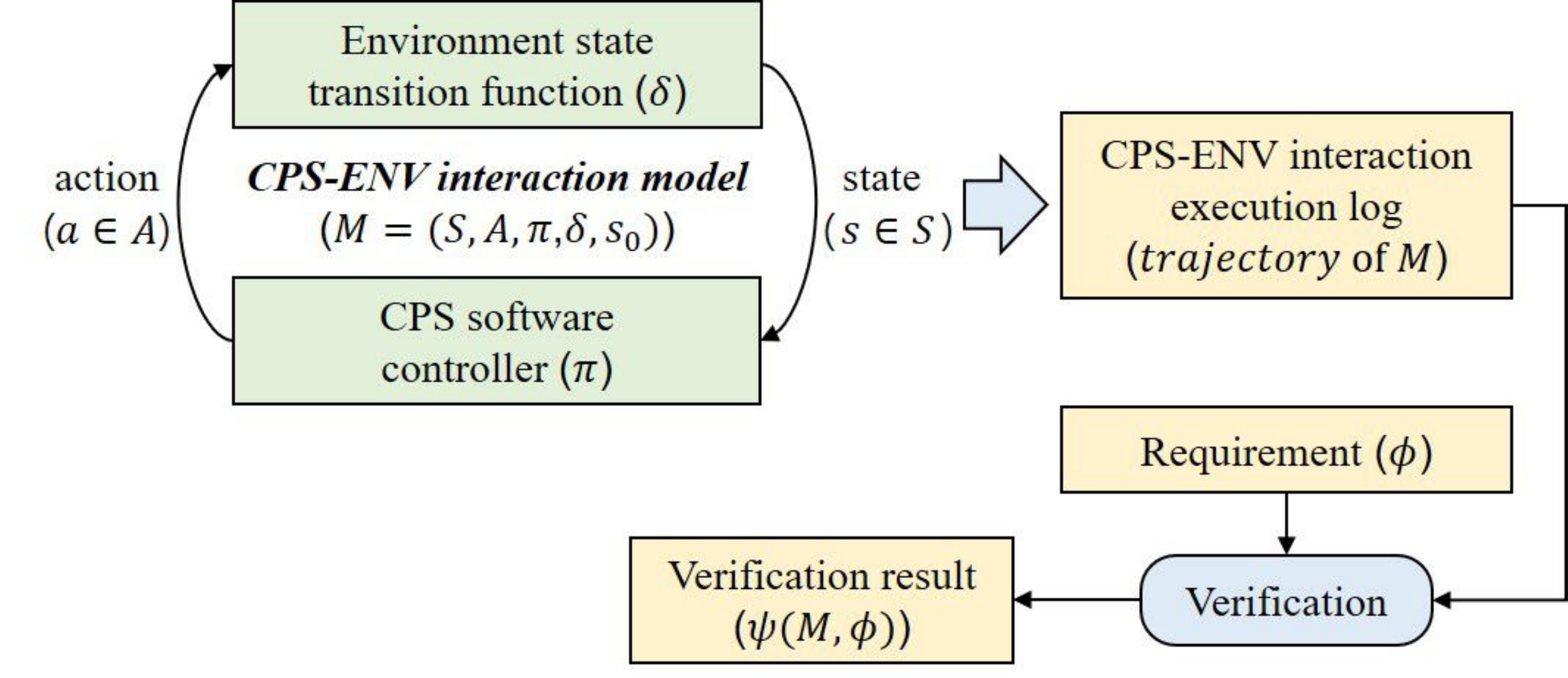}
    \caption{Formal framework for CPS goal verification\label{fig:formalism}}
\end{figure}

Since a trajectory of a \textit{CPS-ENV interaction model} concisely captures the sequential interaction between the CPS under analysis and its environment, one can easily verify whether CPS goals are achieved or not by analyzing the trajectory. \figurename~\ref{fig:formalism} visualizes how a \textit{CPS-ENV interaction model} is used for simulation-based CPS goal verification. Specifically, let $\phi$ be a requirement that precisely specifies a goal under verification. The achievement of $\phi$ is quantifiable. For a \textit{CPS-ENV interaction model} $M$, the verification result of $\phi$ for $M$, denoted by $\psi(M, \phi)$, is computed by evaluating the achievement of $\phi$ on the trajectory of $M$. 
Depending on the type of $\phi$, the value of $\psi(M, \phi)$ can be Boolean (expressing the success or failure of a requirement with clear-cut criteria) or Float (expressing the measurement of an evaluation metric of $\phi$). For example, one of the evaluation metrics of the lane-keeping requirement is the distance the vehicle is away from the center of the lane. As a result of the verification of the lane-keeping goal, the average or maximum distance from the center is computed.

\subsection{Problem Statement}\label{sec:problem-statement}
The problem of virtual environment model generation for simulation-based CPS goal verification is to find an accurate virtual environment model that can replace the real environment of the CPS goal under verification while maintaining the same level of verification accuracy.
Specifically, for the same CPS under analysis, let a \textit{CPS-ENV interaction model} $M_r = (S, A, \pi, \delta_r, s_0)$ representing the interaction between the CPS and its real environment (in FOT) and another model $M_v = (S, A, \pi, \delta_v, s_0)$ representing the interaction between the same CPS and its virtual environment (in simulations). Notice that we have the same $S$, $A$, $\pi$, and $s_o$ for both $M_r$ and $M_v$ since they are about the same CPS\footnote{Note that $S$ can be the same for $M_r$ and $M_v$ because it is a set of \textit{observable} states from the perspective of the CPS under analysis.\label{foot:formalism2}}, whereas $\delta_r$ and $\delta_v$ are different since they represent how the corresponding environments react to the actions performed by the CPS.
For a requirement $\phi$, we aim to have $\delta_v$ that minimizes the difference between $\psi(M_r, \phi)$ and $\psi(M_v, \phi)$. Therefore, the problem of virtual environment model generation for CPS goal verification is to find $\delta_v$ such that $|\psi(M_r, \phi) - \psi(M_v, \phi)|$ is the minimum. 

The virtual environment model generation problem has three major challenges. 
First, the number of possible states and actions is often very large, making it infeasible to build a virtual environment model (i.e., represented by a transition function $\delta_v: S \times A \to S$) by exhaustively analyzing individual states and actions. 
Second, since the virtual environment model continuously interacts with the CPS under analysis in a closed-loop, even a small difference between the virtual and real environments can significantly differ in verification results as it accumulates over time, the so-called compounding error problem. This means that simply having a transition function $\delta_v$ that mimics the behavior of $\delta_r$ in terms of individual input and output pairs, without considering the accumulation of errors for sequential inputs, is not enough.
Third, generating $\delta_v$ should not be as expensive as using many FOTs; otherwise, there is no point in using simulation-based CPS goal verification. Recall that manually crafting virtual environment models in a high-fidelity simulator requires a lot of expertise, which takes longer than doing FOTs many times for having statistically significant verification results. Therefore, a practical approach should generate an accurate virtual environment model efficiently and automatically.

To address the challenges mentioned above, we suggest leveraging IL to automatically generate virtual environment models from only a small amount of data. The data is the partial trajectory of $M_r$, which can be collected from a few FOTs for the CPS under test in its real application environment.
Since IL can efficiently extract how experts make sequential actions for given states from a limited amount of demonstrations while minimizing the compounding errors, it is expected to be an excellent match to our problem. Therefore for our problem, IL will extract $\delta_v$, instead of $\pi$ (which is the original goal of IL), that can best reproduce given trajectories of $M_r$ (i.e., FOT logs). Generated $\delta_v$ may differ depending on the amount of the trajectory, so we analyze it in the experiment.

\section{Environment Imitation}\label{sec:approach}

This section provides \textit{ENVI}, a novel approach to the problem of environment model generation for CPS goal verification, defined in Section \ref{sec:problemDef}. 
We solve the problem by using IL to automatically infer a virtual environment state transition function from the log recorded during the interaction between the CPS under test and its application environment. In this context, the real application environment is considered an ``expert," and the FOT log demonstrates the expert.

\begin{figure*}[ht]
    \centering
    \includegraphics[angle=0, width=\textwidth]{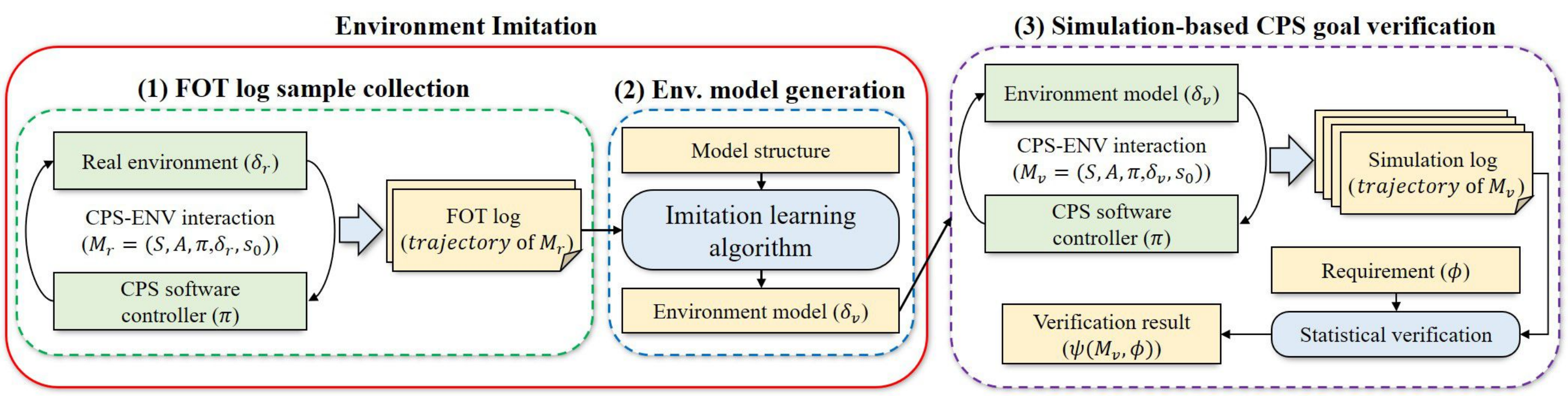}
    \caption{\textit{Environment Imitation} process and simulation-based CPS goal verification\label{fig:overallApproach}}
\end{figure*}

\figurename~\ref{fig:overallApproach} shows the overview of the environment model generation and simulation-based CPS goal verification process using our approach. It is composed of three main stages: (1) FOT log collection for model generation, (2) environment model generation using an IL algorithm, and (3) CPS goal verification using the generated environment model. 
In the first stage, engineers collect FOT logs of a CPS controller under analysis $\pi$ deployed in its real application environment. The interaction between the CPS and the real environment is abstracted as $M_r$, including the unknown $\delta_r$. 
The trajectory of $M_r$ recorded in the logs is then used by IL algorithms in the second stage to generate a virtual environment model $\delta_v$ that imitates $\delta_r$ automatically.
In the last stage, the simulation of $\pi$ in the virtual environment described by $\delta_v$ is performed to generate simulation logs as many as needed for statistical verification. 
As a result, engineers can statistically verify to what extent a requirement $\phi$ of the CPS is satisfied using only a few FOT logs. In the following subsections, we explain each of the main steps in detail with the example introduced in Section \ref{sec:motivatingEx}.

\subsection{FOT Log Collection}\label{subsec:fieldExp}

The first stage of \textit{ENVI} is to collect the interaction data between the CPS controller and its real environment, which will be used as the ``demonstrations'' of imitation learning to generate the virtual environment later. For a CPS-ENV interaction model $M_r=(S, A, \pi, \delta_r, s_0)$ defined in Section~\ref{sec:problemDef}, the interaction data collected over time $T$ can be represented as the trajectory of $M_r$ over $T$ steps, i.e., $\langle (s_0, a_0), (s_1, a_1), \dots, (s_T, a_T)\rangle$ where $s_{t+1}=\delta_r(s_t, a_t)$ and $a_{t}=\pi(s_t)$ for $t\in \{0,1,\dots,T-1\}$. The trajectory can be easily collected from an FOT, since it is common to record the interaction between the CPS controller and its real environment as an FOT log~\cite{xu2019big}. For example, the lane-keeping system records time-series data of the distances the vehicle deviated from the center of the lane $d_t$ and the steering angles $a_t$ over $t=0,1,\dots,T$ during an FOT.

In practice, the trajectory of the same $M_r$ is not necessarily the same due to the uncertainty of the real environment, such as the non-uniform surface friction. Therefore, it is recommended to collect a few FOT logs for the same $M_r$. Since the virtual environment model generated by imitation learning will mimic the given trajectories as much as possible, the uncertainty of the real environment recorded in the trajectories will also be imitated. Section~\ref{sec:caseStudy} will investigate to what extent virtual environment models generated by \textit{ENVI} can accurately mimic the real environment in terms of CPS goal verification when the size of the given FOT logs varies.

\subsection{Environment Model Generation}\label{subsec:environmentImitation}
The second stage of \textit{ENVI} is to generate a virtual environment model from the collected FOT logs using an IL algorithm. It consists of two steps: (1) define the environment model structure and (2) run an IL algorithm to generate a trained model.

\subsubsection{Defining Environment Model Structure}\label{sec:approach-model-structure}
We implement an environment model as a neural network to leverage imitation learning. Before training the environment model, users define the neural network structure. 

\begin{figure}
    \centering
    \includegraphics[width=\linewidth]{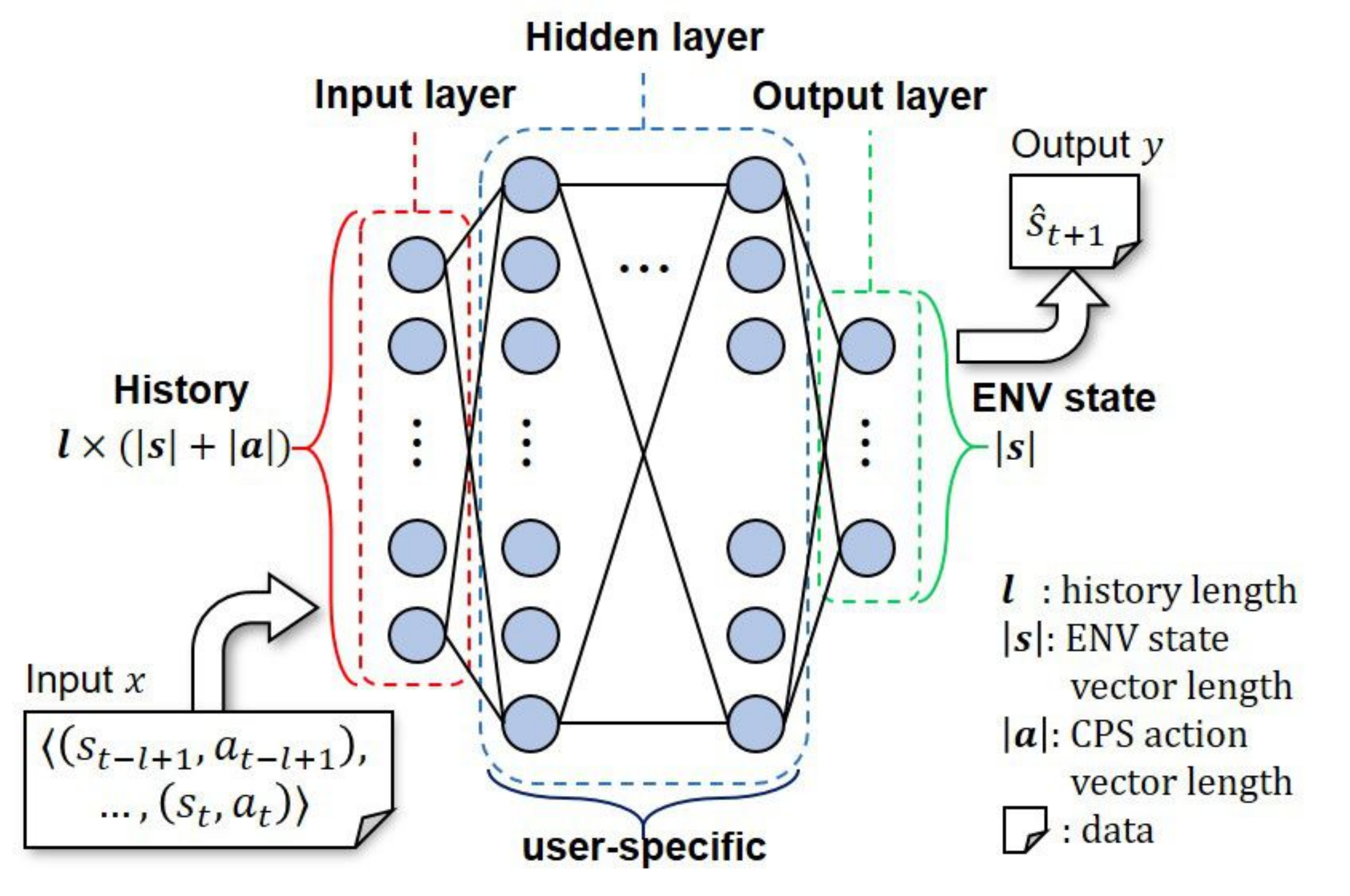}
    \caption{The environment model structure}
    \label{fig:modelStructure}
\end{figure}

The virtual environment model structure is based on the environmental state transition function $\delta: S\times A \to S$ defined in Section~\ref{sec:problemDef}. It assumes that the ideal (real) environment generates the next state $s_{t+1}\in S$ by taking the current environment state $s_t\in S$ and the current CPS action $a_t\in A$ only, meaning that $(s_t, a_t)$ is sufficient to determine $s_{t+1}$ in the ideal environment at time $t$. However, in practice, $s$ may not include sufficient information since it is observed by the sensors of the CPS under verification and the sensors have limited sensing capabilities. 
To solve this issue, we extend $\delta$ for virtual environment models as $\delta_v: (S\times A)^l \to S$ where $l$ is the length of the state-action pairs required to predict the next state. This means that $\delta_v$ uses $\langle (s_{t-l+1}, a_{t-l+1}), \dots, (s_t, a_t) \rangle$ to predict $s_{t+1}$. Notice that $\delta_v$ is equal to $\delta$ when $l=1$. 
To account for the extension of $\delta$, we also extend the CPS-ENV interaction model $M=(S, A, \pi, \delta, s_0)$ to $M_v=(S, A, \pi, \delta_v, \sigma_0)$ where $\sigma_0  = \langle (s_0, a_0), \dots, (s_{l-1}, a_{l-1}) \rangle$ is a partial trajectory of $M_r$ over $l$ steps starting from $s_0$. Intuitively speaking, $\sigma_0$ is the initial input for $\delta_v$ similar to $s_0$ (and $a_0 = \pi(s_0)$) for $\delta$.

Based on the extended definition of $\delta_v$, the structure of $\delta_v$ is shown in \figurename~\ref{fig:modelStructure}. The input and output of $\delta_v$ are $\langle (s_{t-l+1}, a_{t-l+1}), \dots, (s_t, a_t) \rangle$ and $s_{t+1}$, respectively, as defined above. 
Recall that an environmental state $s$ and a CPS action $a$ can be vectors in general; let $|x|$ be the length of a vector $x$. Then, the number of input neurons of the neural network is $l\times (|s|+|a|)$, and the number of output neurons is $|s|$. 

Defining the environment model structure involves two manual tasks. 
The first task is to choose a proper value for the history length $l$. If the value of $l$ increases, more information can be captured in environmental states while the cost of training and executing $\delta_v$ increases. Therefore, it is important to balance the amount of information and the cost of computation. For example, one can visualize the FOT log and see if there are any cyclic patterns in the sequence of environmental states.
The second task is to design the hidden layers of $\delta_v$. The hidden layers specify how the output variables are calculated from the input variables, so-called forward propagation. The design of hidden layers is specific to a domain, but general guidelines of the neural network design exist for practitioners~\cite{hagan1997neural, rafiq2001neural, schilling2018deep}.

\subsubsection{Environment Model Training using IL Algorithms}
Once the structure of $\delta_v$ is determined, we can train $\delta_v$ using an IL algorithm with a proper set of training data $D = \{(X_1, Y_1), \dots, (X_n, Y_n) \}$, where $n$ is the number of FOT logs, $X_i$ is the sequence of inputs collected from $i$-th FOT log and $Y_i$ is the corresponding sequence of outputs (i.e., the expected value of $\delta_v(x_j)$ is $y_j$ for all $j\in \{ 1,\dots,|X_i| \}$ and $|X_i| = |Y_i|$ for $i\in \{1,\dots,n\}$). 
Since $x\in X$ is an $l$-length sequence of state-action pairs, we can generate $D$ from an FOT log using a sliding window of length $l$. Specifically, for an FOT log $\langle (s_0, a_0), \dots, (s_T, a_T) \rangle$, $x_j=\langle (s_j, a_j), \dots, (s_{l-j+1}, a_{l-j+1}) \rangle$ for $j\in \{0, \dots, T-l+1\}$. 

In the following subsections, we explain how each of the representative IL algorithms, i.e., BC, GAIL, and the combination of BC and GAIL, can be used for training $\delta_v$. 

\paragraph{\textbf{Using BC}}
As described in Section \ref{sec:background:bc}, BC trains an environment model $\delta_v$ using supervised learning. Pairs of the input and output of the real environment recorded in FOT logs are given to $\delta_v$ as training data, and $\delta_v$ is trained to learn the real environment state transition shown in the training data. 

Specifically, the BC algorithm (whose pseudocode is shown in Algorithm~\ref{algo:bc}) takes as input a randomly initialized environment model $\delta_v$ and a training dataset $D$; it returns an environment model $\delta_v$ trained using $D$.

\begin{algorithm}
    \SetAlgoLined
    
    \SetKwInOut{Input}{Input}
    \SetKwInOut{Output}{Output}
    
    \Input{ENV model (randomly initialized) $\delta_v$, \newline Training data $D=\{(X_1,Y_1), \dots, (X_n, Y_n)\}$ 
    }
    \Output{ENV model (trained) $\delta_v$}
    
    \While{$\mathit{not (stoping\_condition)}$}{\label{alg:BC:while:start}
        \ForEach{$(X, Y) \in D$}{\label{alg:BC:foreach:start}
            Sequence of model outputs $Y' \gets \delta_v(X)$\label{alg:BC:predict}\\  
	        Float $\mathit{loss}_\mathit{BC}\gets \mathit{getLoss}(Y, Y')$\label{alg:BC:loss}\\
        	$\delta_v\gets \mathit{update}(\delta_v, \mathit{loss}_\mathit{BC})$\label{alg:BC:update}\\
        }\label{alg:BC:foreach:end}
    }\label{alg:BC:while:end}
    \textbf{return} $\delta_v$\label{alg:BC:return}\\
    \caption{\textit{ENVI} BC algorithm\label{algo:bc}}
\end{algorithm}

The algorithm iteratively trains $\delta_v$ using $D$ until a stopping condition (e.g., a fixed number of iterations or convergence of the model's loss) is met (lines~\ref{alg:BC:while:start}--\ref{alg:BC:while:end}). 
For each $(X, Y)\in D$, the algorithm repeats the following (lines~\ref{alg:BC:foreach:start}--\ref{alg:BC:foreach:end}): 
(1) executing $\delta_v$ on $X$ to predict a sequence of outputs $Y'$ (line~\ref{alg:BC:predict}),
(2) calculating the training loss $\mathit{loss_{BC}}$ based on the difference between $Y'$ and $Y$ (line~\ref{alg:BC:loss}), and
(3) updating $\delta_v$ to minimize $\mathit{loss_{BC}}$ (line~\ref{alg:BC:update}). 
The algorithm ends by returning $\delta_v$ (line~\ref{alg:BC:return}).

Algorithm~\ref{algo:bc} is intuitive and easy to implement. In addition, the model's loss converges fast because it is a supervised learning approach. However, if the training data does not fully cover the input space or is biased, the model may not accurately imitate the real environment.

\paragraph{\textbf{Using GAIL}}
As described in Section~\ref{sec:background:gail}, GAIL iteratively trains not only $\delta_v$ but also the discriminator $\zeta$ that evaluates $\delta_v$ in terms of the CPS controller $\pi$. Specifically, for a state $s$, $\zeta$ evaluates $\delta_v$ with respect to $\delta_r$ (captured by $D$) by comparing $\delta_v(s,\pi(s))$ and $\delta_r(s,\pi(s))$. To do this, $\zeta$ is trained using $D$ by supervised learning\footnote{The structure of $\zeta$ is similar to $\delta_v$, but the input of $\zeta$ is $(s, \delta_v(s,\pi(s))$ and the output of $\zeta$ is a reward value $r$.}, and $\delta_v$ is trained using the evaluation results of $\zeta$. %

Algorithm~\ref{algo:gail} shows the pseudocode of GAIL. Similar to Algorithm~\ref{algo:bc}, it takes as input a randomly initialized environment model $\delta_v$ and a training dataset $D = (X, Y)$; however, it additionally takes as input a randomly initialized discriminator $\zeta$ and the CPS controller under analysis $\pi$. It returns a trained virtual environment model $\delta_v$.

\begin{algorithm}
    \SetAlgoLined
    
    \SetKwInOut{Input}{Input}
    \SetKwInOut{Output}{Output}
    \SetKw{iterations}{iterations}
    
    \Input{ENV model (randomly initialized) $\delta_v$, 
    \newline Discriminator (randomly initialized) $\zeta$, 
    \newline Function of CPS decision-making logic $\pi$, 
    \newline Training data $D=\{(X_1,Y_1), \dots, (X_n, Y_n)\}$     
    }
    \Output{ENV model (trained) $\delta_v$}
    
    \While{$\mathit{not (stoping\_condition)}$}{\label{alg:GAIL:while:start}
        \ForEach{$(X, Y) \in D$}{\label{alg:GAIL:foreach:start}
            \tcp{Discriminator training}
            Sequence of model outputs $Y'\gets \delta_v(X)$\label{alg:GAIL:predict}\\
            Float $\mathit{loss}_d \gets \mathit{getDisLoss}(\zeta, X, Y, Y')$\label{alg:GAIL:disLoss}\\
            $\zeta\gets \mathit{update}(\zeta, \mathit{loss}_d)$\label{alg:GAIL:updateDis}
            \BlankLine

            \tcp{Environment model training}
            Sequence of model rewards $R\gets \emptyset$\label{alg:GAIL:InitR}\\
            Model input $x' \gets X[0]$\label{alg:GAIL:InitInput}\\
            \For{$|X|-1$}{\label{alg:GAIL:simul:start}
                Model output $y'\gets \delta_v(x')$\label{alg:GAIL:simul:predict}\\
                Reward $r \gets \zeta(x', y')$\label{alg:GAIL:simul:reward}\\
                $R\gets \mathit{append}(R, r)$\label{alg:GAIL:simul:append}\\
                CPS action $a\gets \pi(y')$\label{alg:GAIL:simul:action}\\
                $x' \gets \mathit{updateInput}(x', y', a)$\label{alg:GAIL:simul:sliding}\\
            }\label{alg:GAIL:simul:end}
            Float $\mathit{loss}_\mathit{GAIL} \gets \mathit{aggregate}(R)$\label{alg:GAIL:lossEnv}\\
            $\delta_v\gets \mathit{update}(\delta_v, \mathit{loss}_\mathit{GAIL})$\label{alg:GAIL:updateEnv}\\
        }\label{alg:GAIL:foreach:end}
    }\label{alg:GAIL:while:end}
    \textbf{return} $\delta_v$\label{alg:GAIL:return}\\
    \caption{\textit{ENVI} GAIL algorithm}
    \label{algo:gail}
\end{algorithm}

The algorithm iteratively trains both $\delta_v$ and $\zeta$ using $D$ and $\pi$ until a stopping condition is met (lines~\ref{alg:GAIL:while:start}--\ref{alg:GAIL:while:end}). 
To train $\zeta$, for each $(X,Y)\in D$ (lines~\ref{alg:GAIL:foreach:start}--\ref{alg:GAIL:foreach:end}), the algorithm executes $\delta_v$ on $X$ to predict a sequence of outputs $Y'$ (line~\ref{alg:GAIL:predict}), calculates the discriminator loss $\mathit{loss}_d$ indicating how well $\zeta$ can distinguish $Y$ and $Y'$ for $X$ (line~\ref{alg:GAIL:disLoss}), and updates $\zeta$ using $\mathit{loss}_d$ (line~\ref{alg:GAIL:updateDis}). 
Once $\zeta$ is updated, the algorithm trains $\delta_v$ using $\zeta$ and $\pi$ (lines~\ref{alg:GAIL:InitR}--\ref{alg:GAIL:updateEnv}). Specifically, the algorithm initializes a sequence of rewards $R$ (line~\ref{alg:GAIL:InitR}) and a model input $x'$ (line~\ref{alg:GAIL:InitInput}), collects $r\in R$ for each $x'$ using $\delta_v$, $\pi$, and $\zeta$ (lines~\ref{alg:GAIL:simul:start}--\ref{alg:GAIL:simul:end}), calculates the environment model loss $loss_{GAIL}$ by aggregating $R$ (line~\ref{alg:GAIL:lossEnv}), and updates $\delta_v$ using $loss_{GAIL}$ (line~\ref{alg:GAIL:updateEnv}).
To collect $r\in R$ for each $x'$ (lines~\ref{alg:GAIL:simul:start}--\ref{alg:GAIL:simul:end}), the algorithm executes $\delta_v$ on $x'$ to predict an output $y'$ (line~\ref{alg:GAIL:simul:predict}), executes $\zeta$ on $x'$ and $y'$ to get a reward $r$ (line~\ref{alg:GAIL:simul:reward}), appends $r$ at the end of $R$ (line~\ref{alg:GAIL:simul:append}), executes $\pi$ on $y'$ to decide a CPS action $a$ (line~\ref{alg:GAIL:simul:action}), and updates $x'=\langle (s_1, a_1), (s_2, a_2) \dots, (s_l, a_l) \rangle$ as $x'=\langle (s_2, a_2) \dots, (s_l, a_l), (y', a) \rangle$ by removing $(s_1, a_1)$ and appending $(y', a)$ (line~\ref{alg:GAIL:simul:sliding}).
The algorithm ends by returning $\delta_v$ (line~\ref{alg:GAIL:return}).

Notice that, to train $\delta_v$, GAIL uses the input-output pair $(x', y')$ simulated by $\pi$ and $\zeta$, in addition to the real input-output pair $(x, y)$ in $D$. This is why it is known to work well even with a small amount of training data~\cite{ho2016generative, jena2020augmenting}. However, the algorithm is more complex to implement than BC, and the environment model converges slowly or sometimes fails to converge depending on hyperparameter values.

\paragraph{\textbf{Using BC and GAIL together}}
Notice that BC trains $\delta_v$ using the training data only, but GAIL trains $\delta_v$ using the simulated data as well; BC and GAIL can be combined to use both training and simulated data without algorithmic conflict. This idea is suggested by \citet{ho2016generative} to improve learning performance, and \citet{jena2020augmenting} later implemented the idea as an algorithm BCxGAIL. 

The BCxGAIL algorithm is the same as GAIL in terms of its input and output, and it also trains both $\delta_v$ and $\zeta$ similar to GAIL. In particular, $\zeta$ is updated as the same as in GAIL. However, $\delta_v$ is updated using both $\mathit{loss}_\mathit{BC}$ (line 4 in Algorithm~\ref{algo:bc}) and $\mathit{loss}_\mathit{GAIL}$ (line 15 in Algorithm~\ref{algo:gail}). By doing so, BCxGAIL can converge fast (similar to BC) with a small amount of training data (similar to GAIL).

\subsection{Simulation-based CPS Goal Verification}\label{subsec:usecases}
Using the virtual environment model $\delta_v$ generated from the previous stage, an engineer can statistically verify if the CPS controller $\pi$ under analysis satisfies a goal $\phi$ (i.e., compute $\psi(M_v, \phi)$) through many simulations of $M_v = (S, A, \pi, \delta_v, \sigma_0)$. 

To simulate $M_v$, the initialization data $\sigma_0$ should be given. Since $\sigma_0$ is the partial trajectory of $M_r$ over $l$ steps, the engineer should conduct partial FOTs over $l$ steps to get $\sigma_0$. 
Notice that acquiring $\sigma_0$ is much cheaper than having full FOTs for FOT-based CPS goal verification since $l$ is much shorter than $T$ (i.e., the full FOT duration).
The engineer then run $M_v$ as many times as needed for statistical verification\footnote{This is because $\delta_v$ can be non-deterministic and the same $\sigma_0$ can lead to different simulation results.}. For example, to verify if a vehicle equipped with a lane-keeping system under development is not more than $\SI{1}{\meter}$ away from the center of the lane, engineers simulate the lane-keeping system several times with the generated environment model. The engineers then analyze the distance farthest from the center of the lane in each simulation and verify whether the requirement is statistically satisfied. 

\begin{figure}
    \centering
    \includegraphics[width=\linewidth]{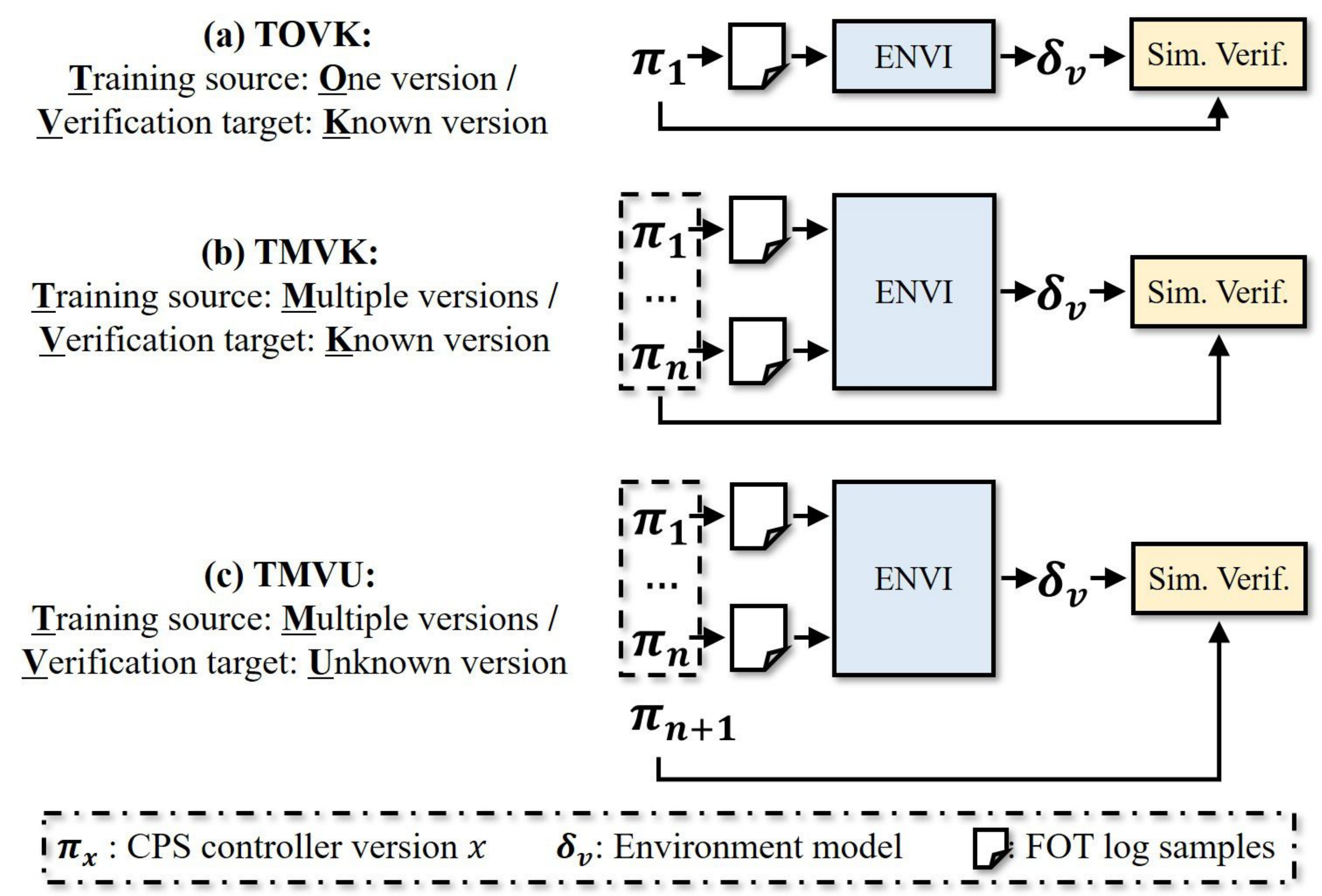}
    \caption{Use cases of simulation-based verification using \textit{ENVI}\label{fig:usecases}}
\end{figure}

In practice, it is common to develop multiple versions of the same CPS controller, for example, developed sequentially during its evolutionary development~\cite{basden1991evolutionary, helps2012comprehensive, sirjani2021towards}. Let us consider a lane-keeping system controller implemented with a configuration parameter indicating the minimum degree of steering for lane-keeping. Then, one can develop a new version of the lane-keeping system by changing the parameter value based on the CPS goal verification results of its previous versions. 
In such an evolutionary development process, for the verification of the new version, we can consider different use cases depending on which version of the FOT logs is used to generate the environment model. Specifically, we can consider three different use cases:

\paragraph{\textbf{Case 1}: One version is used for training, and verification is performed on the same version as training}
This is the basic use case, shown in \figurename~\ref{fig:usecases} (a). For example, for the verification of the first version of the lane-keeping system controller, some FOT logs of that version must be collected since there are no previous versions (and their FOT logs). Since Training involves One version and Verification is for the Known version, we refer to this case \textit{TOVK}. 

\paragraph{\textbf{Case 2}: Multiple versions are used for training, and verification is performed on one of the versions used for training}
Multiple versions of the CPS controller can be used for training, as shown in \figurename~\ref{fig:usecases} (b). For example, when there are different sets of FOT logs collected by previously developed versions of the lane-keeping system in addition to the FOT logs collected by the new version, all the logs associated with different parameter values can be used together to generate a single environment model. This allows us to best utilize all FOT logs for virtual environment model generation. Since Training involves Multiple versions and Verification is for one of the Known versions, we refer to this case \textit{TMVK}. 

\paragraph{\textbf{Case 3}: Multiple versions are used for training, and verification is performed on a new version that has never been used for training}
As shown in \figurename~\ref{fig:usecases} (c), this is similar to the TMVK use case, but without using FOT logs collected by the new version. In other words, only the previously collected FOT logs are used for the verification of the new version. This allows us to significantly reduce the cost of new FOTs for the new version for CPS goal verification. Since Training involves Multiple versions and Verification is for an Unknown version, we refer to this case \textit{TMVU}. 

\section{Case Study}\label{sec:caseStudy}
This section provides a case study to evaluate the applicability of our approach in various use cases introduced in Section~\ref{subsec:usecases}. Specifically, we first investigate the accuracy of CPS goal verification results when \textit{ENVI} is used for a single CPS controller version (i.e., the TOVK use case). We then analyze if \textit{ENVI} can efficiently generate a single environment model that can be used for the CPS goal verification of multiple CPS controller versions (i.e., the TMVK use case). Last but not least, we also investigate if the single environment model can be used for the CPS goal verification of a new CPS controller version that has never been used for training (i.e., the TMVU use case). To summarize, we answer the following research questions:
\begin{enumerate}[\bf RQ1:]
    \item Can \textit{ENVI} generate a virtual environment model that can replace the real environment in the CPS goal verification for a single CPS controller version? (\textbf{\textit{TOVK}})
    \item Can \textit{ENVI} generate a virtual environment model that can replace the real environment in the CPS goal verification for multiple CPS controller versions? (\textbf{\textit{TMVK}})
    \item Can \textit{ENVI} generate a virtual environment model that can replace the real environment in the CPS goal verification for a new CPS controller version? (\textbf{\textit{TMVU}})
\end{enumerate}

\subsection{Subject CPS}\label{subsec:subjectCPS}

\begin{figure}
    \centering
    \includegraphics[width=0.8\linewidth]{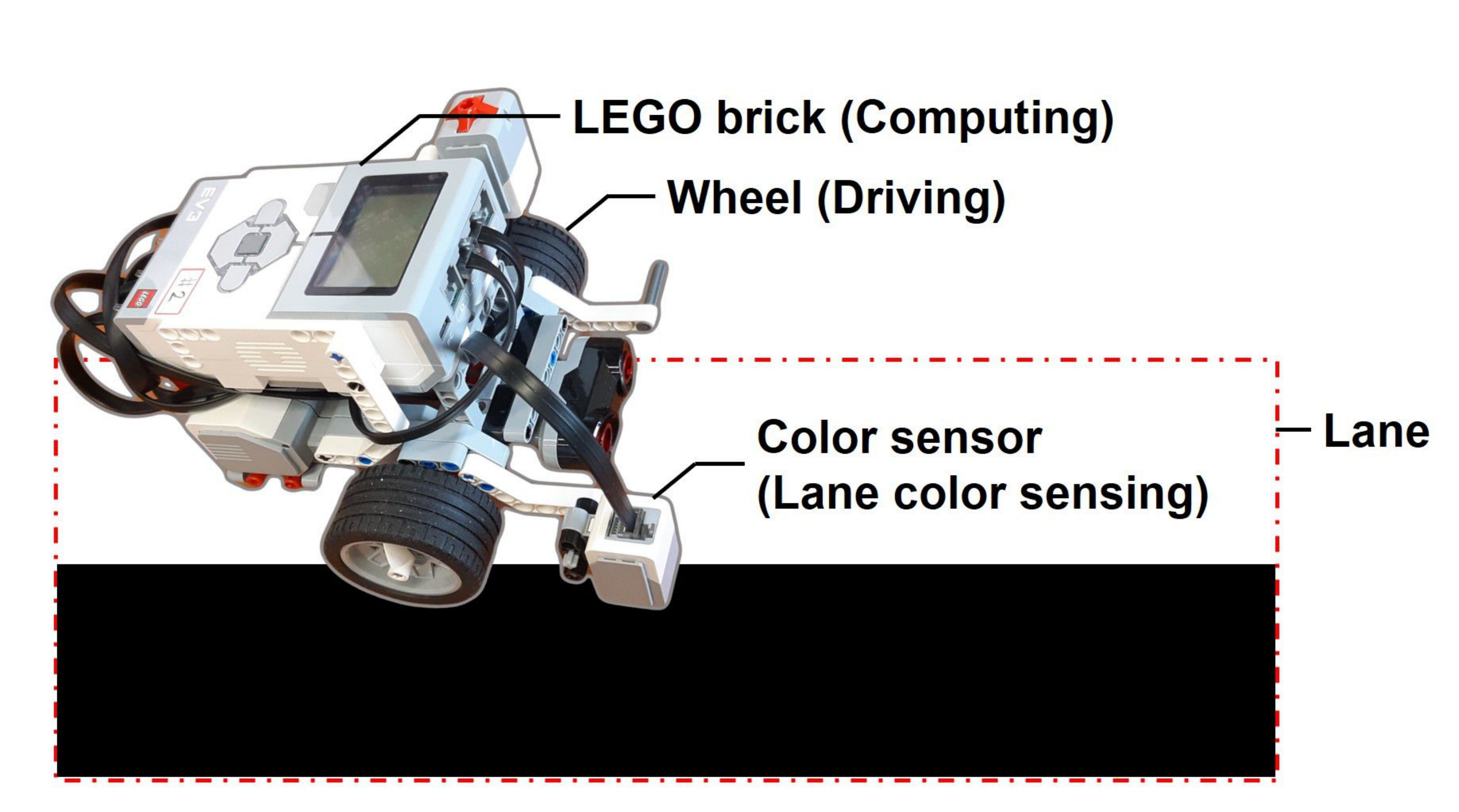}
    \caption{Case study subject CPS: a LEGO-lized autonomous vehicle\label{fig:robot}}
\end{figure}

To answer the research questions in the context of a real CPS development process, we implement a simplified autonomous vehicle equipped with a lane-keeping system. We utilize an open physical experimental environment~\cite{shin2021platooning} that abstracts an autonomous vehicle as a programmable LEGO robot and a road as a white and black paper lane, as shown in \figurename~\ref{fig:robot}. The goal of the lane-keeping system is to keep the center of the lane, indicated by the border between white and black areas while driving, so we aim to verify the goal achievement of the lane-keeping system (e.g., how smoothly it drives following the lane center). Similar to many other CPS, the LEGO-lized autonomous vehicle comprises three parts: sensor, controller, and actuator. A sensor (e.g., a color sensor) gives data observing the CPS environment to a controller. A controller (e.g., a Python program in a LEGO brick) controls actuators (e.g., a motor of a wheel) that make CPS act.

As for the controller, we aim to consider multiple versions of the same lane-keeping system and compare them using simulation-based CPS goal verification to find the best one that allows the ego vehicle to drive the smoothest along the center of the lane. 
To do this, we develop a template of rule-based lane-keeping system logic and instantiate it into multiple versions of the same lane-keeping system with different parameter values. 
Algorithm~\ref{algo:laneKeeping} shows the template logic with a configurable parameter $x$ indicating the degree of rotation; the algorithm takes as input a color $c$ (range from 0 meaning the darkest to 100 meaning the brightest) from the color sensor and returns an angle $a$ for the rotation motor. Positive/negative angle means turning right/left, respectively. The algorithm simply turns right if the value of $c$ is greater than 50 (i.e., the color is darker than gray) and turns left if the value of $c$ is less than 50 (i.e., the color is lighter than gray); otherwise (i.e., the color is exact gray), the algorithm goes straight. The parameter value of $x$ determines the degree of turning right and left. 
We consider five different parameter values of $x$, i.e., from $10\degree$ to $50\degree$ in steps of $10\degree$ in our case study.

\begin{algorithm}
    \SetAlgoLined
    
    \SetKwInOut{Config}{Config.}
    \SetKwInOut{Input}{Input}
    \SetKwInOut{Output}{Output}
   
    \Config{Positive float of unit rotation degree $x$}
    \Input{Float of lane color value $c$}
    \Output{Float of rotation angle value $a$}

    Float $gray\gets 50$\\
    
    \If{$c>gray$}{
        $a\gets x$ \hspace{17pt}\tcp{Turn right}
    }
    \ElseIf{$c<gray$}{
        $a\gets -x$ \hspace{10pt}\tcp{Turn left}
    }
    \Else{
        $a\gets 0$ \hspace{17pt}\tcp{Go straight}
    }
    \textbf{return} $a$
    
    \caption{Lane-keeping system controller logic\label{algo:laneKeeping}}
\end{algorithm}

\begin{figure}
    \centering
    \includegraphics[width=\linewidth]{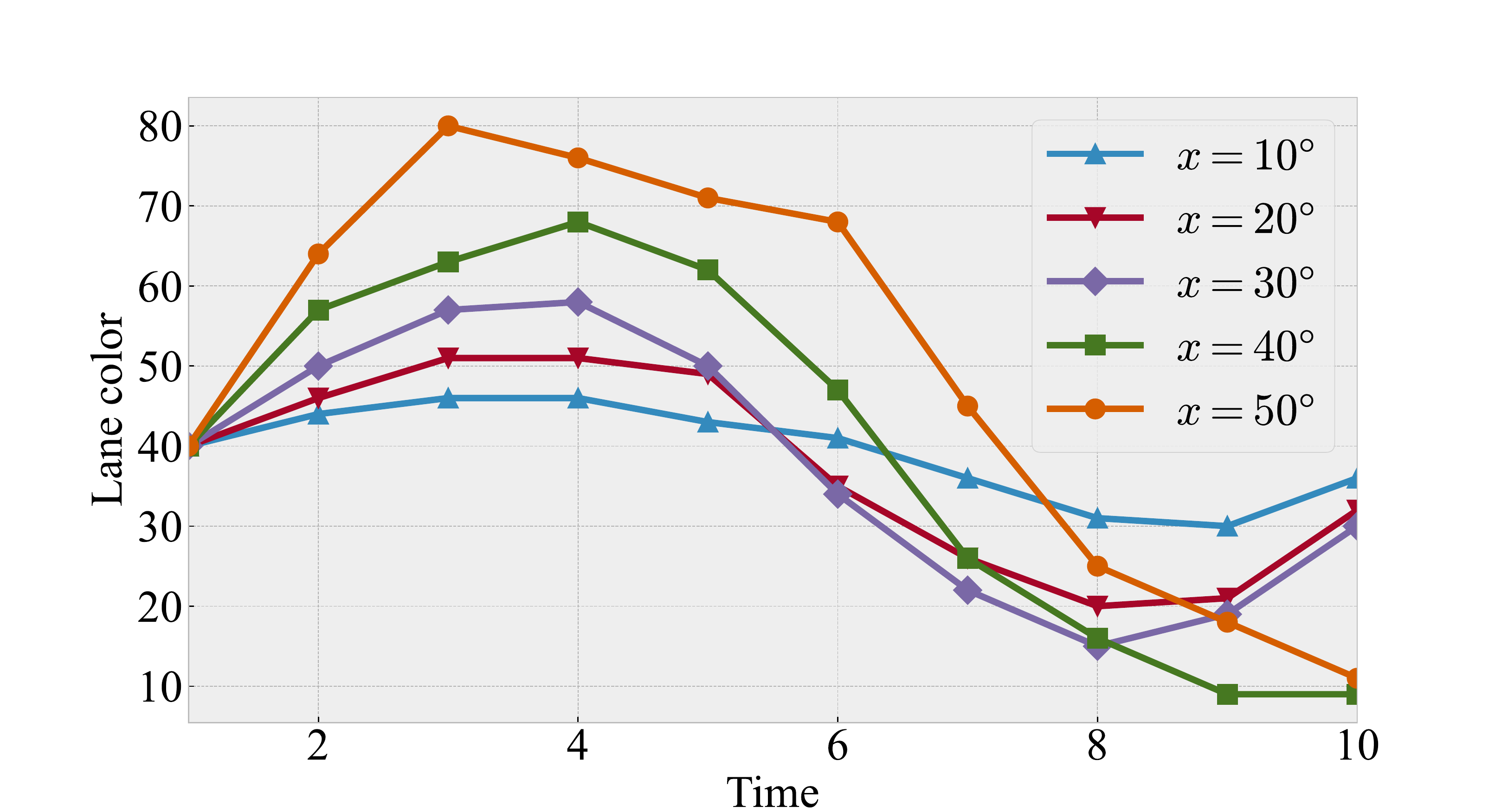}
    \caption{Different interactions between the CPS and its real environment depending on different CPS controller versions}
    \label{fig:FOTlog}
\end{figure}

Although the algorithm simplifies the logic of lane-keeping systems with a configuration parameter $x$, making a parameterized controller and optimizing the controller's configuration are common in practice~\cite{tao2008novel, david2012three}. In addition, engineers experience that changing the configuration changes the CPS behavior in the real environment. \figurename~\ref{fig:FOTlog} shows the partial FOT logs of the lane-keeping system with different $x$ values used in our case study; we can see how the interaction between the lane-keeping system and the real environment varies depending on the value of $x$.

Based on Algorithm~\ref{algo:laneKeeping}, we implement five different CPS controllers to cover the three use cases (i.e., TOVK, TMVK, and TMVU) described in Section~\ref{subsec:usecases}. Specifically, we follow an evolutionary development~\cite{helps2012comprehensive} scenario where 
(1) a CPS controller with $x=30\degree$ is developed, and its goal achievement is verified first (i.e., TOVK),
(2) two versions with $x=10\degree$, and $x=50\degree$ are additionally developed, and an environment model is generated using FOT logs of $x=10\degree$, $x=30\degree$, and $x=50\degree$ together and is used to verify each developed version (i.e., TMVK), and
(3) two more versions with $x=20\degree$ and $x=40\degree$ are additionally developed, and their goal achievements are verified using the previously generated virtual environment model without using any FOT logs for $x=20\degree$ and $x=40\degree$ (i.e., TMVU).

\begin{figure}
    \centering
    \includegraphics[width=1.0\linewidth]{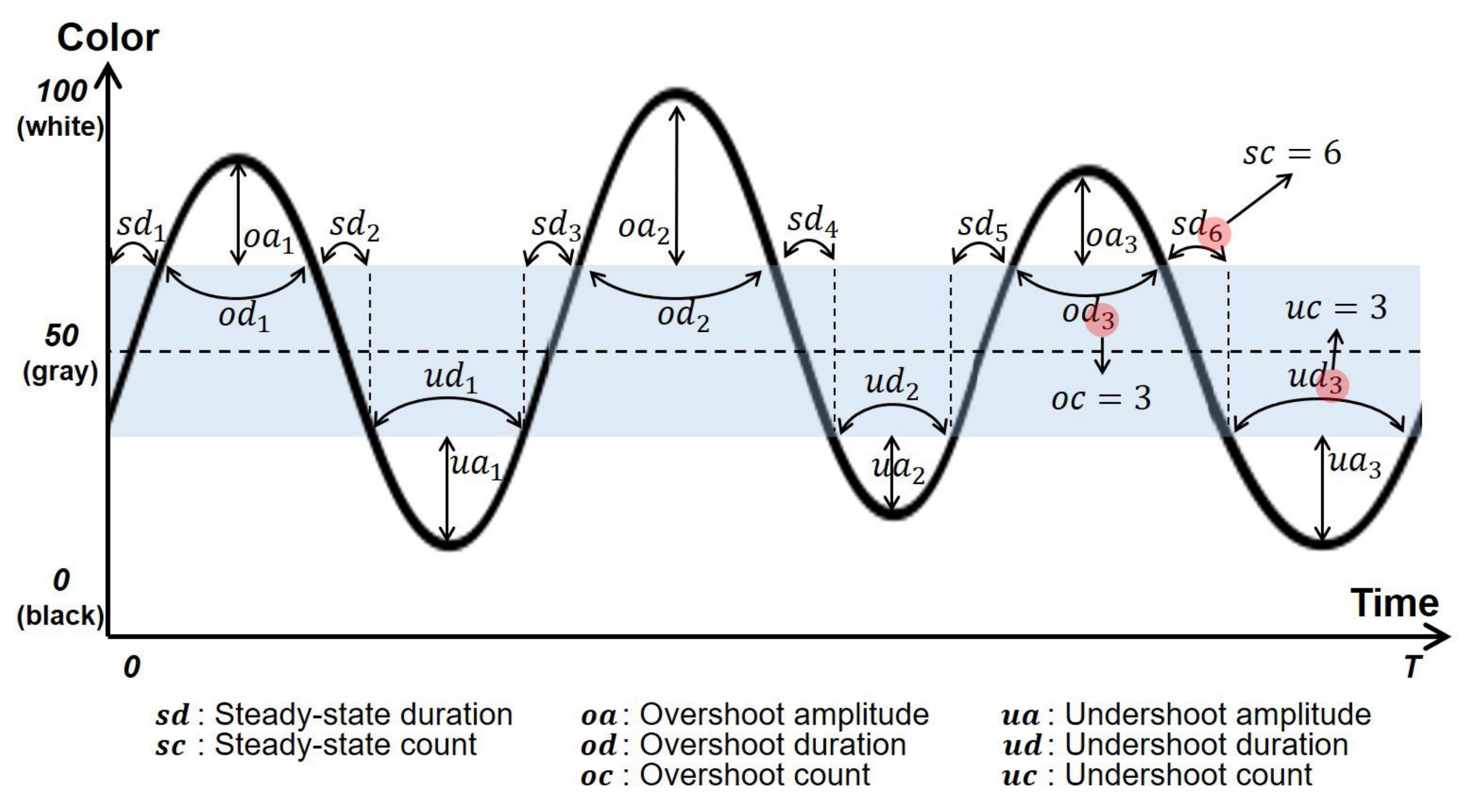}
    \caption{Driving quality metrics in a lane-keeping system log\label{fig:laneKeepingLog}}
\end{figure}

To assess the goal achievement of the different CPS controllers (i.e., how smoothly the vehicle drives following the lane), we define multiple driving performance metrics by investigating driving traces collected during our preliminary experiments~\cite{cherrett2001extracting, shin2021platooning}. Specifically, given time length $T$, eight driving quality metrics are defined as follows (visualized in \figurename~\ref{fig:laneKeepingLog}): 
\begin{compactenum}[1)]
    \item number of steady-state $sc$: indicating how many times the vehicle stays in the lane center thresholds
    \item total steady-state duration $\sum_{i=1}^{sc}sd_i$: indicating how long the vehicle stays in the lane center thresholds
    \item number of overshooting $oc$: indicating how many times the vehicle overshoots the upper threshold of the lane center
    \item sum of overshooting amplitudes $\sum_{i=1}^{oc}oa_i$: indicating how much the vehicle overshoots
    \item total overshooting duration $\sum_{i=1}^{oc}od_i$: indicating how long the vehicle overshoots
    \item number of undershooting $uc$: indicating how many times the vehicle undershoots the lower threshold of the lane center 
    \item sum of undershooting amplitudes $\sum_{i=1}^{uc}ua_i$: indicating how much the vehicle undershoots
    \item total undershooting duration $\sum_{i=1}^{uc}ud_i$: indicating how long the vehicle undershoots
\end{compactenum}
It is straightforward that the smaller the metrics about the overshooting and undershooting (i.e., metric 3--8), the better. In addition, if the vehicle does not deviate from the lane center, the steady-state continues uninterrupted, and its duration becomes $T$.
Therefore, at the ideal case (e.g., driving exactly on the lane center), the first metric $sc=1$, the second metric $\sum_{i=1}^{sc}sd_i=T$, and the other metrics are all $0$.

\subsection{\textit{ENVI} Experimental Setup}\label{subsec:enviSetting}
As described in Section~\ref{sec:approach}, the CPS goal verification using \textit{ENVI} follows three main stages: (1) FOT log collection, (2) environment model generation, and (3) simulation-based CPS goal verification. In the following subsections, we explain our experimental setup for each stage in detail.

\subsubsection{FOT Log Collection}\label{sec:eval-logs}
For each of the five CPS controller versions, we conduct 30 FOTs of the simplified autonomous vehicle and collect 30 logs to capture how the vehicle interacts with the real environment. At each time $t$, the following information is recorded in the logs: (1) a lane color value $c_t$ as an environmental state observed by the vehicle's color sensor and (2) a steering angle $a_t$ as a CPS action decided by the vehicle's controller. Therefore, an FOT log is a sequence of state-action pairs $\langle(c_0, a_0), (c_1, a_1), \dots, (c_T, a_T)\rangle$ where $T$ is the FOT duration. According to the vehicle's hardware spec, it records 25 state-action pairs per one second. Since a sequence of 25 state-action pairs is enough to observe the behavior of a CPS controller, we set $T$ to 25 (i.e., one FOT log is collected by one second).

\subsubsection{Environment Model Generation}\label{sec:eval-modelgen}
To investigate the impact of using different IL algorithms, we generate different environment models using BC, GAIL, and BCxGAIL. We implement the algorithms in PyTorch~\cite{paszke2019pytorch}. BC uses the ADAM optimizer~\cite{kingma2014adam} to update environment models. Since GAIL needs a policy gradient algorithm to update models~\cite{ho2016generative}, we use a state-of-the-art Proximal Policy Optimization (PPO) algorithm~\cite{schulman2017proximal}. As for the hyperparameters of the IL algorithms, we best use default values from the original papers~\cite{schulman2017proximal, ho2016generative}. Table~\ref{tab:setting} shows the hyperparameter values used in our evaluation. 

\begin{table}
\centering
\caption{Hyperparameter values for IL algorithms}
\footnotesize
\label{tab:setting}
\begin{tabular}{llr}
\toprule
                   Algorithm   & Hyperparameter              & Value   \\
\midrule
\multirow{2}{*}{BC}   & Epoch                       & 300     \\ 
                      & Learning rate               & 0.00005 \\
\midrule
\multirow{8}{*}{GAIL} & Epoch                       & 300     \\ 
                      & Model learning rate         & 0.00005 \\ 
                      & PPO num. policy iteration        & 10      \\  
                      & Discriminator learning rate & 0.01    \\ 
                      & PPO num. discriminator iteration & 10      \\ 
                      & PPO reward discount $\gamma$                   & 0.99    \\
                      & PPO GAE parameter $\lambda$                  & 0.95    \\
                      & PPO clipping $\epsilon$                  & 0.2     \\
\bottomrule
\end{tabular}
\end{table}

As for the model structure, we set the length of history $l$ as 10, meaning that the input of a virtual environment model is a 20-dimensional vector (i.e., a sequence of 10 state-action pairs). We use a simple design for hidden layers for both the virtual environment model and discriminator. Tables~\ref{tab:envModelLayer} and \ref{tab:discModelLayer} summarize the structures of virtual environment model and discriminator, respectively. 

\begin{table}
\centering
\caption{Environment model structure}
\footnotesize
\label{tab:envModelLayer}
\begin{tabular}{llr}
\toprule
\# & layer                 & \# output units \\ \midrule
   & input                 & 20             \\ 
1  & fully connected layer & 256                        \\ 
2  & tanh & 256                        \\
3  & fully connected layer & 256                        \\ 
4  & tanh & 256                        \\ 
5  & fully connected layer & 1                          \\ 
6  & tanh & 1                         \\ 
\bottomrule
\end{tabular}
\end{table}

\begin{table}
\centering
\caption{Discriminator structure}
\footnotesize
\label{tab:discModelLayer}
\begin{tabular}{llr}
\toprule
\# & layer                 & \# output units \\ \midrule
   & input                 & 21                        \\ 
1  & fully connected layer & 256                        \\ 
2  & ReLU & 256                        \\
3  & fully connected layer & 256                        \\ 
4  & ReLU & 256                        \\ 
5  & fully connected layer & 1                          \\ 
6  & Sigmoid & 1                          \\ 
\bottomrule
\end{tabular}
\end{table}

As for the model training, to better understand the training data efficiency of each of the IL algorithms, we vary the number of FOT logs to be used and compare the resulting models. Specifically, among the 30 FOT logs collected in Section~\ref{sec:eval-logs}, we randomly select $n$ logs for training a virtual environment model and vary $n$ from 3 to 30 in steps of 3. For all models, normalized state and action values (ranging between $-1$ and $+1$) recorded in the FOT logs are used for training.

\subsubsection{Simulation-based CPS Goal Verification}
To verify the CPS goal achievement, each of the five CPS controller versions is simulated multiple times with the environment models generated by \textit{ENVI}, and the simulation logs are used to assess the degree of CPS goal achievement in terms of the eight driving performance metrics defined in Section~ \ref{subsec:subjectCPS}.

\subsection{CPS Goal Verification Accuracy}\label{subsec:eval-metric}
To evaluate how accurate the simulation-based verification using \textit{ENVI} is with respect to the FOT-based verification using enough FOTs for a set of verification requirements, we measure the similarity between the FOT-based verification and simulation-based verification results. 
Specifically, for a set of CPS goals (requirements) $\Phi$, the CPS goal verification accuracy $\mathit{acc}$ of a virtual environment model $\delta_v$ for the CPS (controller) $\pi_x$ is defined as 
\begin{equation*}
\mathit{acc}(\delta_v, \pi_x) = 1 - \frac{\Sigma_{\phi \in \Phi} |\psi(M_{x,r},\phi)-\psi(M_{x,v},\phi)|}{|\Phi|}	
\end{equation*}
where $M_{x,r} = (S, A, \pi_x, \delta_r, s_0)$ represents the interaction between the CPS under verification (represented by $\pi_x$) and its real environment and $M_{x,v} = (S, A, \pi_x, \delta_v, \sigma_0)$ represents the interaction between the same CPS and $\delta_v$.
As for the set of requirements $\Phi$, we consider eight CPS goals (requirements) $\phi_1, \phi_2, \dots, \phi_8$ based on the eight driving performance metrics defined Section~\ref{subsec:subjectCPS}. Since individual requirements $\phi \in \Phi$ have different ranges, we normalize $\psi(\cdot,\phi)$ to a value between 0 and 1 using the possible minimum and maximum values. As a result, $\mathit{acc}(\delta_v)$ ranges between 0 and 1; the higher its value, the more accurate the virtual environment model. 

To compute $\psi(M_r,\phi)$ for $\phi \in \Phi$, we perform 100 FOTs using our autonomous robot vehicle and collect the FOT logs. Note that these logs are for evaluating the simulation-based verification accuracy, and therefore different from the 30 FOT logs used for training virtual environment models described in Section~\ref{subsec:enviSetting}.

To compute $\psi(M_v,\phi)$ for $\phi \in \Phi$, we perform 100 simulations using a virtual environment model $\delta_v$ generated by \textit{ENVI}. 
To make the FOT-based verification and the simulation-based verification compatible, the same initials $\sigma_0$ must be used. To achieve this, we provide $\delta_v$ with the initial ten pairs of states and actions of each of the 100 FOT logs as $\sigma_0$ for each simulation. 

Notice that $\sigma_0$ is the only real data given to $\delta_v$ to compute $\psi(M_v,\phi)$. From $\delta_v$'s point of view, the CPS under verification $\pi_x$ is black-box and the value of its configuration parameter $x$ is unknown to $\delta_v$, meaning that $\delta_v$ predicts how the real environment continuously interacts with a black-box $\pi_x$ given $\sigma_0$ only.

To account for the randomness in measuring $\psi(M_r,\phi)$ and $\psi(M_v,\phi)$, we repeat the experiment 30 times and report the average.

\subsection{Comparison Baseline}
It is ideal to compare \textit{ENVI} with other existing environment model generation approaches. However, to the best of our knowledge, there is no such approach. Therefore, we make a random environment model for an alternative comparison baseline. The random environment model changes the environmental state randomly regardless of CPS actions. As a result, in addition to the three IL algorithms, we use four different virtual environment model generation approaches (BC, GAIL, BCxGAIL, and Random) and compare them in terms of $\mathit{acc}$.

\subsection{Experiment Results}

\subsubsection{RQ1: \textit{TOVK} Use Case}
RQ1 aims to evaluate whether \textit{ENVI} can generate a virtual environment model for the CPS goal verification of a single CPS controller version. To answer RQ1, we generate a virtual environment model $\delta_{v,30\degree}$ using the FOT logs of the controller version with $x=30\degree$ and measure $\mathit{acc}(\delta_{v,30\degree}, \pi_{30\degree})$ where $\pi_{30\degree}$ indicates the CPS controller version with $x=30\degree$. 

\begin{figure*}[t]
    \centering
    \includegraphics[width=0.7\linewidth]{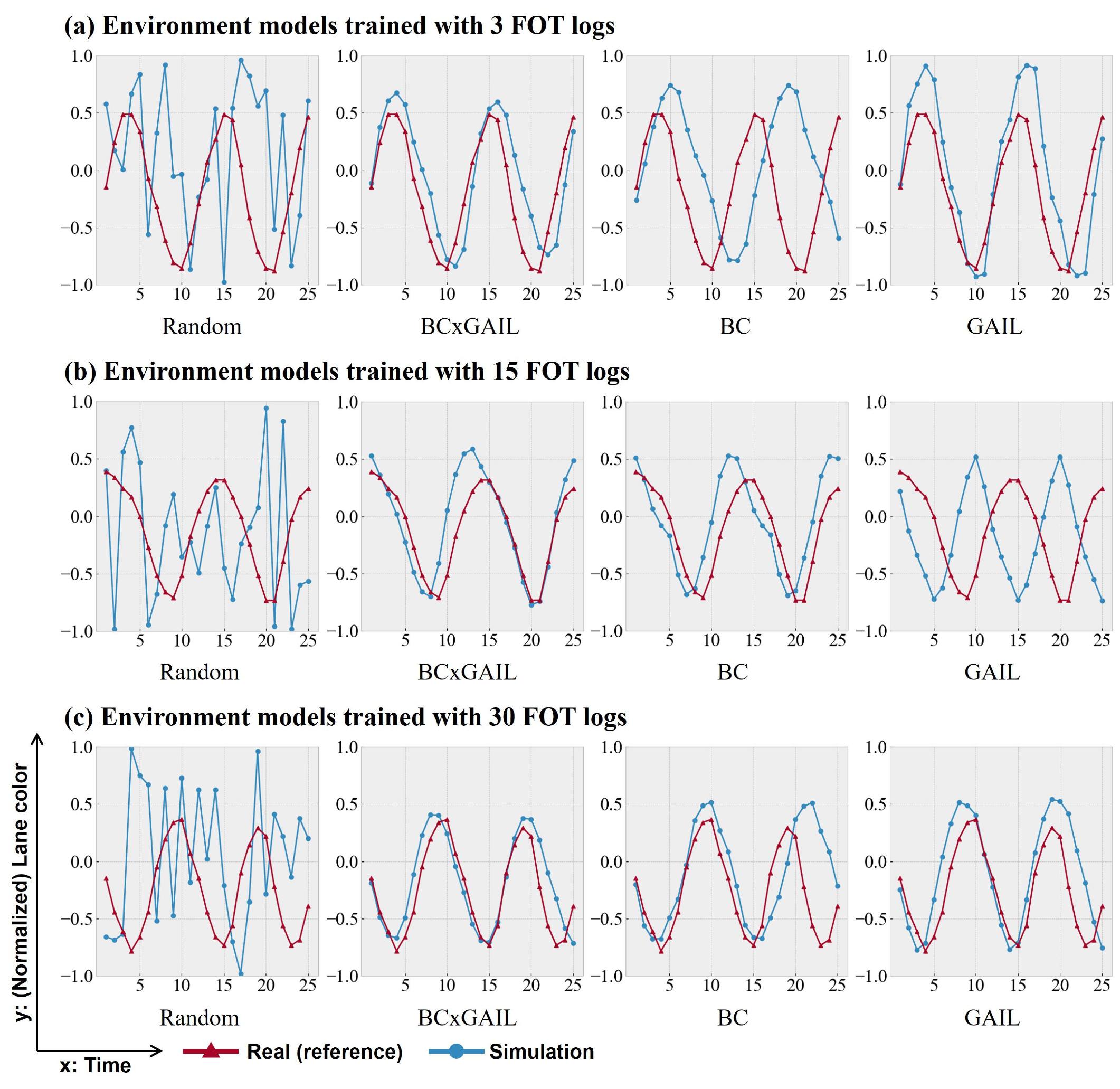}
    \caption{Comparison of real (i.e., FOT) and simulation log data}
    \label{fig:rq1SimulationVis}
\end{figure*}

Before we investigate $\mathit{acc}(\delta_{v,30\degree}, \pi_{30\degree})$ with different configurations (i.e., the different model generation algorithms and the different numbers of FOT logs used for training), we first visualize the behaviors of the real and virtual environments when interacting with the CPS. 
\figurename~\ref{fig:rq1SimulationVis} shows the behaviors of real (red) and virtual (blue) environments in terms of the environmental states (y-axis) generated by the continuous interaction with the lane-keeping system over time (x-axis), when the number of FOT logs used for training is 3 (\figurename~\ref{fig:rq1SimulationVis}(a)), 15 (\figurename~\ref{fig:rq1SimulationVis}(b)), and 30 (\figurename~\ref{fig:rq1SimulationVis}(c)). 
When compared to random, we can see that BC, GAIL, and BCxGAIL generate virtual environment models that can closely mimic the real environment. 
Considering the fact that a slight difference between the virtual and real lane colors at a moment can be accumulated over time due to the closed-loop interaction between the virtual environment model and the lane-keeping system, the visualization shows that all the IL algorithms can learn how the real environment interacts with the lane-keeping system over time without significant errors. 
Moreover, for each of the IL algorithms, the more FOT logs are used, the closer the virtual and real environment models' behaviors are. 
Given the promising visualization results, we continue to investigate the CPS goal verification accuracy below.

\begin{figure}
    \centering
    \includegraphics[width=0.8\linewidth]{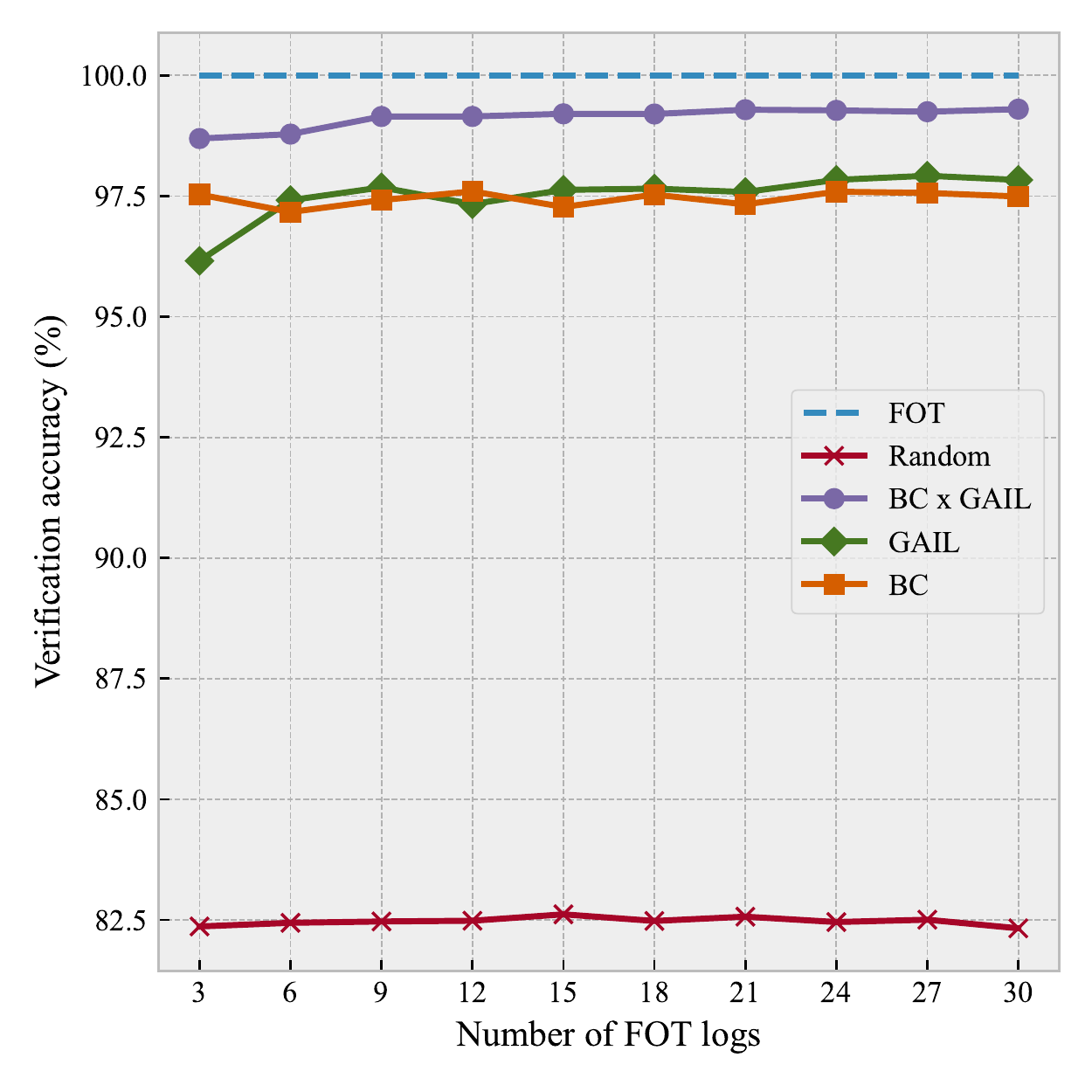}
    \caption{Verification accuracy of environment model generation approaches for TOVK use case\label{fig:tovk}}
\end{figure}

\figurename~\ref{fig:tovk} shows how the CPS goal verification accuracy varies depending on the number of FOT logs used for training the virtual environment model when different model generation algorithms are used. Table~\ref{tab:tovk} additionally provides the accuracy values for representative cases (i.e., when the number of FOT logs is 3, 15, and 30).
Overall, due to the characteristics of the eight driving performance metrics and their normalizations, the random approach's verification accuracy is around 82.5\%. Nevertheless, all environment models generated by \textit{ENVI} achieve higher than 96\% verification accuracy, which is much higher than that of the random approach. This means that the IL algorithms used in \textit{ENVI} are significantly better than the random baseline in terms of generating accurate virtual environment models. 
Regarding the training data efficiency, the verification accuracy only slightly increases as the number of used FOT logs increases, implying that \textit{ENVI} can generate an accurate environment model using even a very small number of FOT logs (e.g., three). 
Comparing the IL algorithms, we can clearly see that BCxGAIL outperforms the others regardless of the number of used FOT logs. This is because the convergence speed and data efficiency of the model training have been complemented by using BC and GAIL algorithms together, as explained in Section \ref{subsec:environmentImitation}. This suggests that engineers can expect the highest model accuracy in this use case through the BCxGAIL algorithm.

\begin{table}
\caption{Verification accuracy results for the TOVK use case. The best accuracy for each number of FOT logs is highlighted in bold.}
\centering
\footnotesize
\label{tab:tovk}
\begin{tabular}{lrr}%
\toprule
Algorithm & Logs & $\mathit{acc}(\delta_{v,30\degree}, \pi_{30\degree})$            \\ 
\midrule
\multicolumn{2}{l}{Random}     -  & 82.47\%        \\ \midrule%
\multirow{3}{*}{BCxGAIL} & 3 & \textbf{98.69}\%  \\ %
                         & 15  & \textbf{99.20}\%  \\ %
                            & 30       & \textbf{99.30}\%  \\ \midrule
\multirow{3}{*}{GAIL}       & 3      & 96.15\% \\ %
                            & 15       & 97.62\% \\ %
                            & 30       & 97.83\% \\ \midrule
\multirow{3}{*}{BC}         & 3      & 97.53\%  \\ %
                            & 15       & 97.27\%  \\ %
                            & 30       & 97.49\%  \\ 
\bottomrule
\end{tabular}
\end{table}

\begin{tcolorbox}
The answer to RQ1 is that \textit{ENVI} can generate an accurate virtual environment model that can replace the real environment in FOTs using only a small number of FOT logs. Among the three IL algorithms used in \textit{ENVI}, BCxGAIL outperforms the others in terms of the CPS goal verification accuracy.
\end{tcolorbox}

\subsubsection{RQ2: \textit{TMVK} Use Case}
RQ2 aims to evaluate whether \textit{ENVI} can generate a virtual environment model for the CPS goal verification of multiple CPS controller versions. To answer RQ2, we measure the verification accuracy of the same virtual environment model for different lane-keeping system controller versions. 
Specifically, we first train $\delta_{v,10\degree|30\degree|50\degree}$ using the FOT logs of three lane-keeping system controller versions (i.e., $x=10\degree$, $x=30\degree$, and $x=50\degree$) and then assess $\mathit{acc}(\delta_{v,10\degree|30\degree|50\degree}, \pi_x)$ for each $x\in\{10\degree, 30\degree, 50\degree\}$.

\begin{figure}
     \centering
     \begin{subfigure}[b]{\linewidth}
         \centering
         \includegraphics[width=0.8\linewidth]{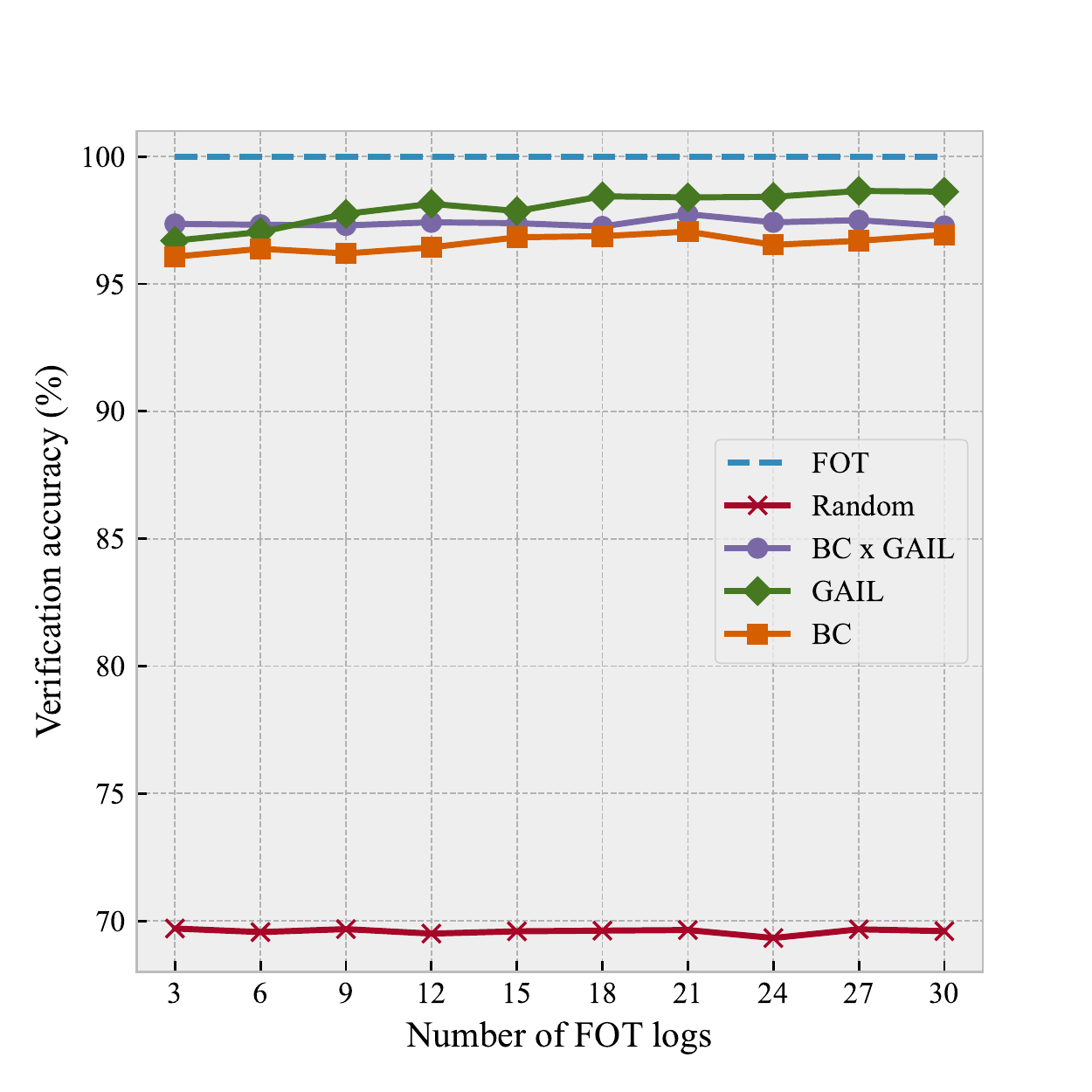}
         \caption{Controller under verification $x=10\degree$}
         \label{fig:tmvk10}
     \end{subfigure}
     \hfill
     \begin{subfigure}[b]{\linewidth}
         \centering
         \includegraphics[width=0.8\linewidth]{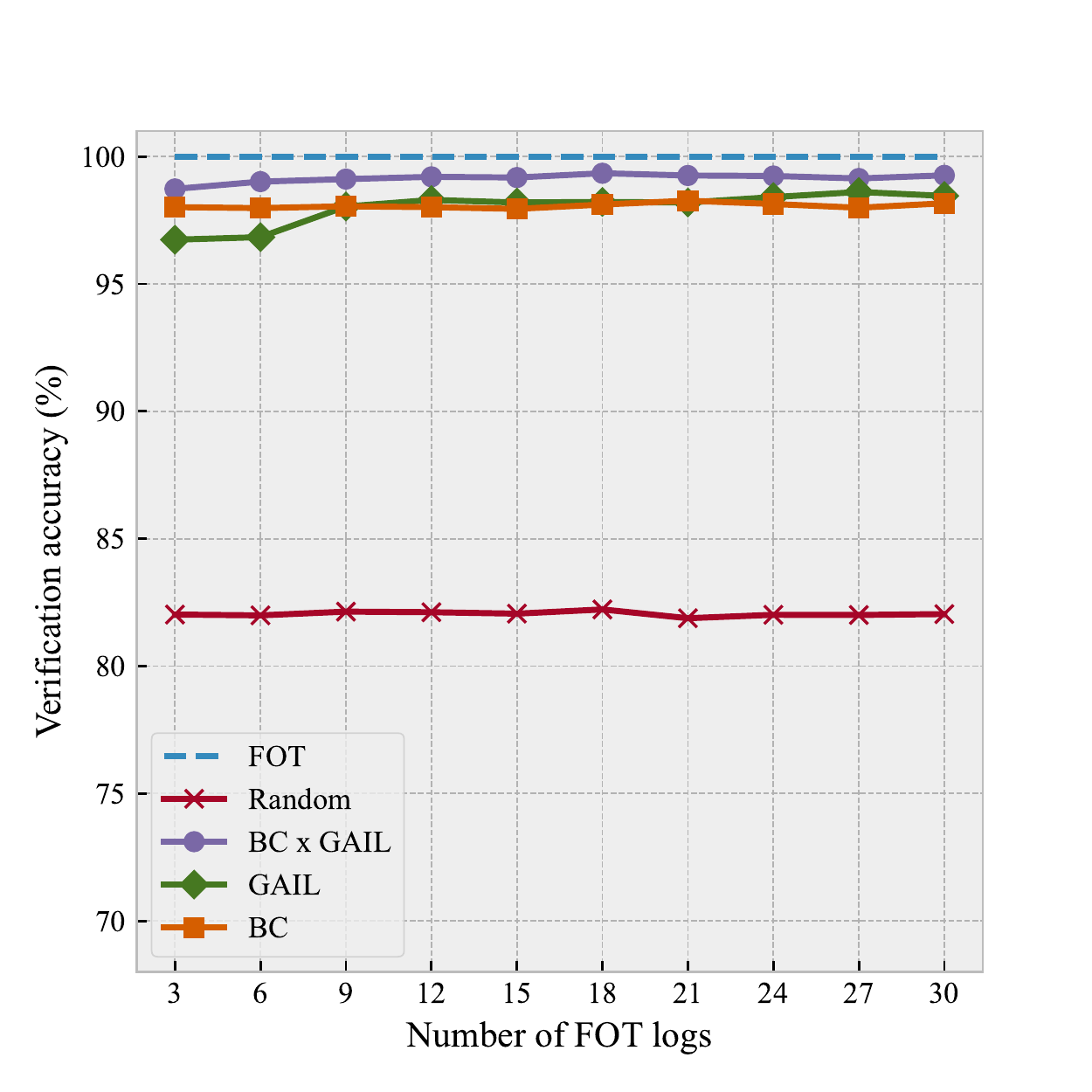}
         \caption{Controller under verification $x=30\degree$}
         \label{fig:tmvk30}
     \end{subfigure}
     \hfill
     \begin{subfigure}[b]{\linewidth}
         \centering
         \includegraphics[width=0.8\linewidth]{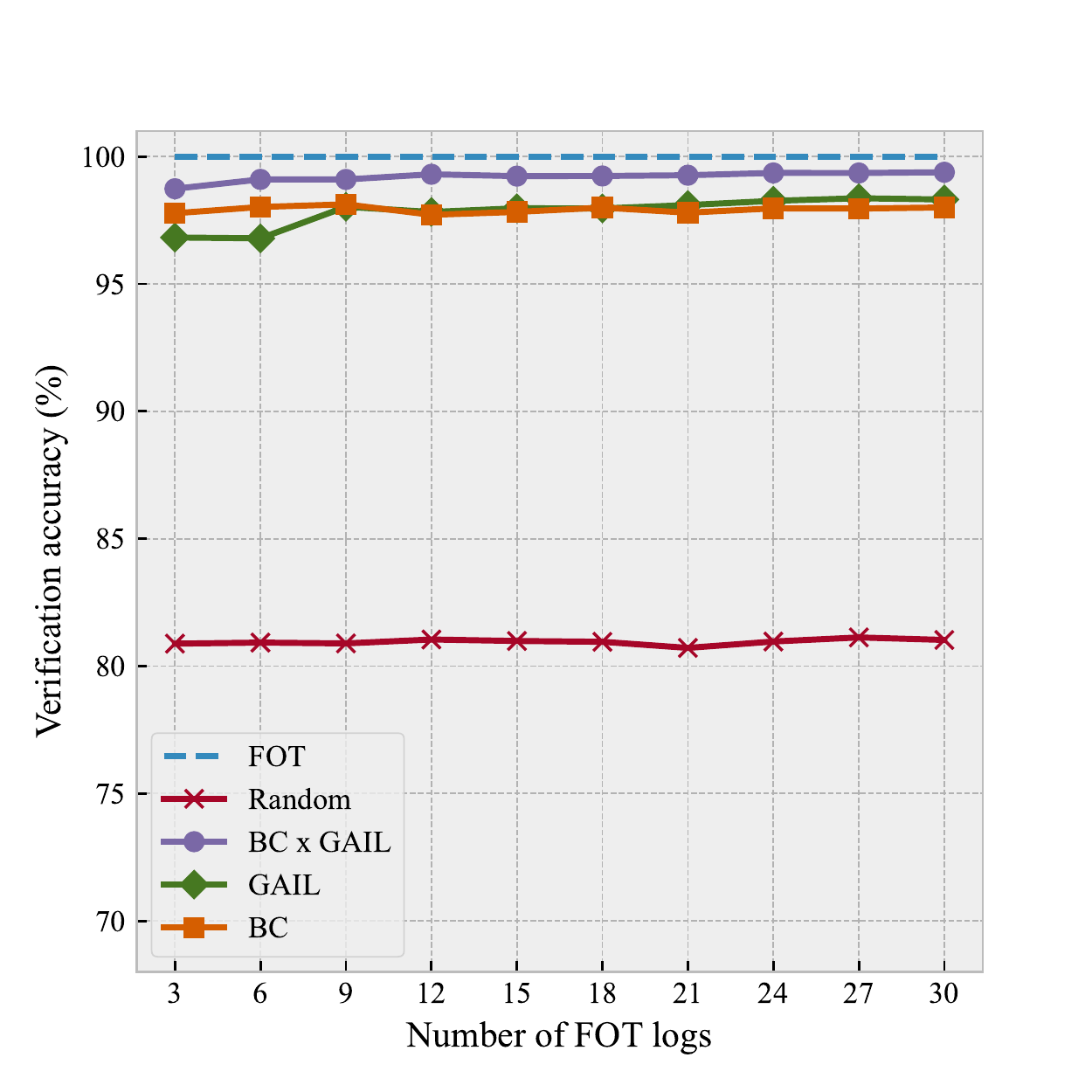}
         \caption{Controller under verification $x=50\degree$}
         \label{fig:tmvk50}
     \end{subfigure}
    \caption{Verification accuracy of environment model generation approaches for TMVK use case}
    \label{fig:tmvk}
\end{figure}

\begin{table}
\caption{Verification accuracy results for the TMVK use case. The best accuracy for each number of FOT logs and for each controller is highlighted in bold.}
\centering
\footnotesize
\label{tab:tmvk}
\begin{tabular}{lrrrrr}
\toprule
& & \multicolumn{3}{c}{$\mathit{acc}(\delta_{v,10\degree|30\degree|50\degree}, \pi_x)$}      &              \\ \cmidrule(r){3-5}
Algorithm & Logs & $x=10\degree$              & $x=30\degree$               & $x=50\degree$      & Avg.             \\ \midrule
Random & - & 69.63\%          & 82.04\%          & 80.96\% & 77.54\%          \\ \midrule
\multirow{3}{*}{BCxGAIL} & 3 & \textbf{97.38}\% & \textbf{98.78}\% & \textbf{98.54}\% & \textbf{98.23}\% \\ 
                         & 15  & 96.95\% & \textbf{99.20}\% & \textbf{99.24}\% & \textbf{98.46}\% \\ 
                            & 30       & 97.47\%          & \textbf{99.33}\% & \textbf{99.34}\% & \textbf{98.71}\%          \\ \midrule
\multirow{3}{*}{GAIL}       & 3      & 96.16\%          & 96.64\%          & 97.25\% & 96.68\%          \\ 
                            & 15       & \textbf{98.21}\%          & 98.32\%          & 97.92\% & 98.15\%          \\ 
                            & 30       & \textbf{98.52}\% & 98.59\%          & 98.80\% & 98.63\%          \\ \midrule
\multirow{3}{*}{BC}         & 3      & 95.77\%          & 97.74\%          & 97.80\% & 97.10\%          \\ 
                            & 15       & 96.14\%          & 97.91\%          & 97.85\% & 97.30\%          \\ 
                            & 30       & 97.17\%          & 98.15\%          & 97.71\% & 97.67\%          \\ \bottomrule
\end{tabular}
\end{table}

\figurename~\ref{fig:tmvk} shows the verification accuracy results depending on the number of training FOT logs for the three different controller versions. Table~\ref{tab:tmvk} provides the accuracy values when the number of FOT logs is 3, 15, and 30. In the table, the best accuracy for each controller version and the number of FOT logs is highlighted in bold. 
Overall, the virtual environment models generated by \textit{ENVI} using the FOT logs of multiple controller versions achieve much higher verification accuracy (at least 95\%) than the random model, even when only a few FOT logs are used for training them. 
Considering the different interaction patterns for different CPS controller versions as shown in \figurename~\ref{fig:FOTlog}, the high verification accuracy indicates that, even with small FOT logs, the IL algorithms can learn how the real environment interacts with different CPS controller versions and generate a single virtual environment model that covers all the different interaction patterns. 
Comparing the IL algorithms, BCxGAIL generates the most accurate environment models using the same number of logs than the other algorithms in general, whereas GAIL sometimes outperforms BCxGAIL when $x=10\degree$ and BC never outperforms the others. This is because GAIL can infer an accurate environment model with small training data (e.g., less than 30) better than BC, as already demonstrated by \citet{ho2016generative}.

\begin{tcolorbox}
The answer to RQ2 is that \textit{ENVI} can generate an accurate virtual environment model that can be shared in the CPS goal verification of different CPS controller versions. Among the IL algorithms used in \textit{ENVI}, BCxGAIL generally outperforms the others in terms of the CPS goal verification accuracy.
\end{tcolorbox}

\subsubsection{RQ3: \textit{TMVU} Use Case}
RQ3 aims to evaluate whether \textit{ENVI} can generate a virtual environment model for the CPS goal verification of a new controller version that has never been used for training the model.
To answer RQ3, we measure the verification accuracy of the virtual environment models generated by multiple controller versions for a new controller version.
Specifically, we first generate $\delta_{v,10\degree|30\degree|50\degree}$ as the same as RQ2 and then assess $\mathit{acc}(\delta_{v,10\degree|30\degree|50\degree}, \pi_x)$ for each new $x\in\{20\degree, 40\degree\}$.

\begin{figure}[t]
    \centering
    \begin{subfigure}[b]{\linewidth}
        \centering
        \includegraphics[width=0.8\linewidth]{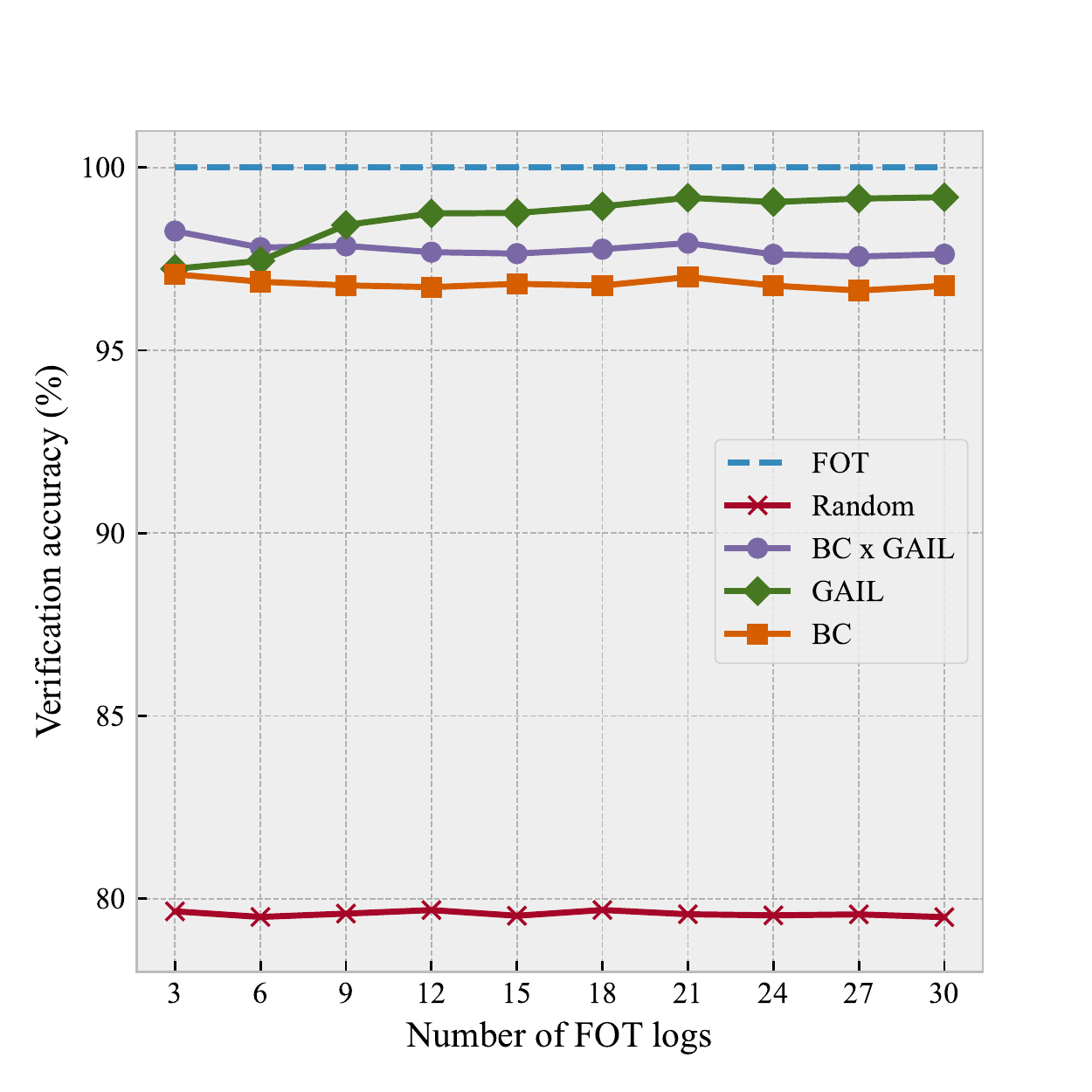}
        \caption{Controller under verification $x=20\degree$}
        \label{fig:tmvu20}
    \end{subfigure}
    \hfill
    \begin{subfigure}[b]{\linewidth}
        \centering
        \includegraphics[width=0.8\linewidth]{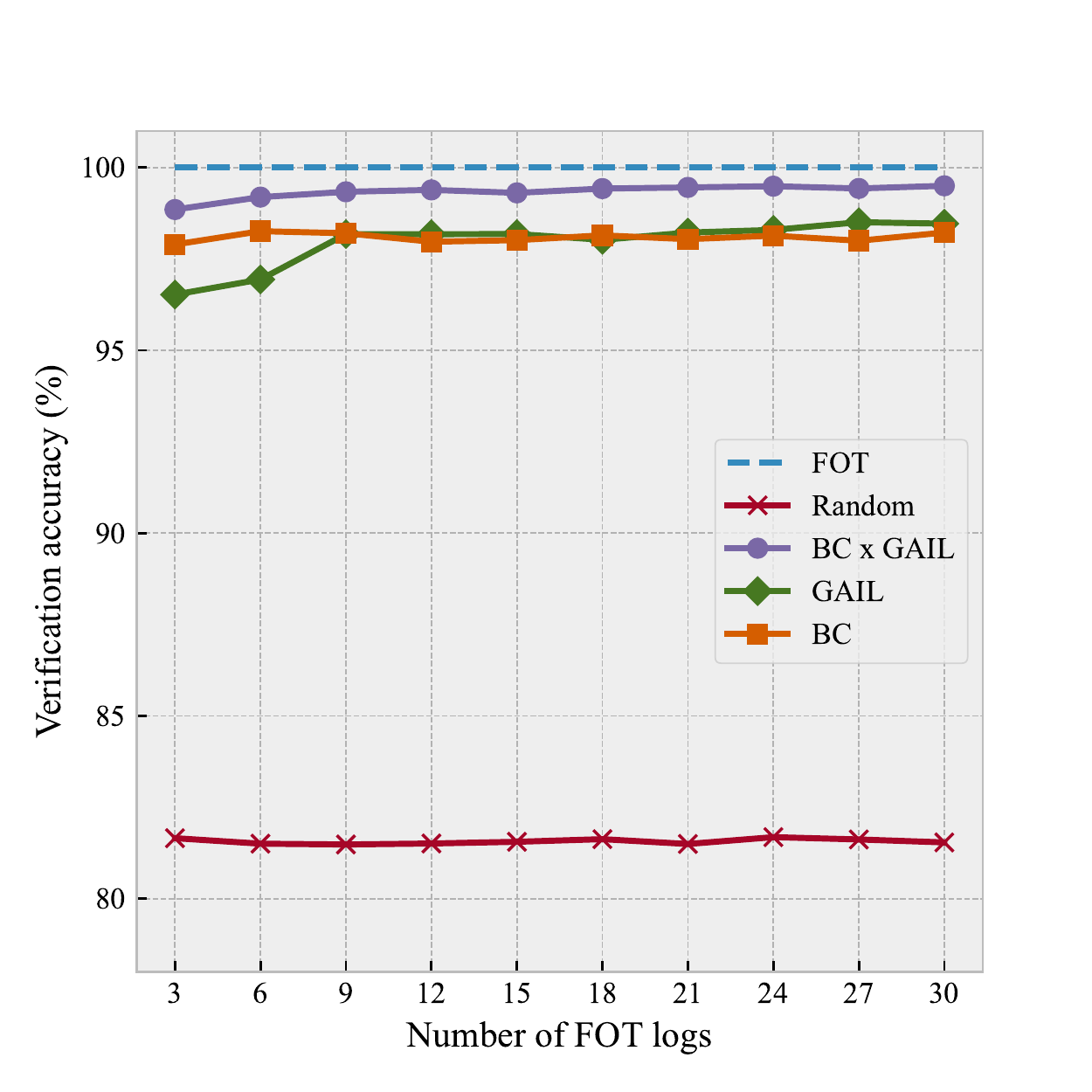}
        \caption{Controller under verification $x=40\degree$}
        \label{fig:tmvu40}
    \end{subfigure}
    \caption{Verification accuracy of environment model generation approaches for TMVU use case.}
    \label{fig:tmvu}
\end{figure}

\begin{table}
\caption{Verification accuracy results for the TMVU use case. The best accuracy for each number of FOT logs and for each controller is highlighted in bold.}
\centering
\footnotesize
\label{tab:tmvu}
\begin{tabular}{lrrrr}
\toprule
& & \multicolumn{2}{c}{$\mathit{acc}(\delta_{v,10\degree|30\degree|50\degree}, \pi_x)$}      &              \\ \cmidrule{3-4}
Algorithm & Logs & $x=20\degree$              & $x=40\degree$                     & avg.             \\ \midrule
Random & -      & 79.63\%          & 81.60\%          & 80.61\%          \\ \midrule
\multirow{3}{*}{BCxGAIL}     & 3     & \textbf{98.44}\% & \textbf{98.76}\% & \textbf{98.60}\% \\ %
                             & 15      & 97.47\%          & \textbf{99.32}\% & 98.40\% \\ %
                             & 30      & 97.51\%          & \textbf{99.53}\% & 98.52\% \\ \midrule
\multirow{3}{*}{GAIL}        & 3     & 97.03\%          & 96.85\%          & 96.94\%          \\ %
                             & 15      & \textbf{98.94}\%        & 98.18\%          & \textbf{98.56}\%          \\ %
                             & 30      & \textbf{99.21}\%        & 98.73\%          & \textbf{98.97}\%          \\ \midrule
\multirow{3}{*}{BC}          & 3     & 97.09\%          & 97.79\%          & 97.44\%          \\ %
                             & 15      & 96.62\%          & 98.18\%          & 97.40\%          \\ %
                             & 30      & 96.71\%          & 98.03\%          & 97.37\%          \\ 
\bottomrule
\end{tabular}
\end{table}

Similar to RQ2, \figurename~\ref{fig:tmvu} and Table~\ref{tab:tmvu} show the verification accuracy results. In all cases, $\delta_{v,10\degree|30\degree|50\degree}$ achieves more than 96\% accuracy, which is much higher than random. 
This means that the virtual environment model generated using the FOT logs of the previously developed CPS controller versions ($x=10\degree$, $x=30\degree$, and $x=50\degree$) can be used for the CPS goal verification for newly developed versions ($x=20\degree$ and $x=40\degree$) with high accuracy. 
This implies that the virtual environment model can learn interaction patterns between the real environment and different CPS controller versions and generalize the patterns to unknown CPS controller versions. 
Therefore, only simulating the new CPS controller versions many times, without much FOT, is required for the CPS goal verification of the new versions if an accurate virtual environment model has been created in the TMVK use case.
This can significantly reduce the cost of the CPS goal verification in practice. 
Regarding the IL algorithms, GAIL and BCxGAIL outperform BC as the same as in RQ2. 
This implies that GAIL and BCxGAIL are recommended for the IL algorithm when \textit{ENVI} is used for the CPS goal verification of new CPS versions.

\begin{tcolorbox}
The answer to RQ3 is that \textit{ENVI} can generate an accurate virtual environment model to verify unknown controller versions that have never been field-tested for training. Regarding the IL algorithms, BCxGAIL and GAIL outperform BC in all cases. 
\end{tcolorbox}

\subsection{Threats to Validity}\label{sec:threat}

In terms of external validity, our case study focused on only a lane-keeping system in a simplified CPS implemented as a LEGO-lized autonomous vehicle and used only one parameter (i.e., the degree of rotations $x$) for representing different versions of software controllers. Although the subject CPS of our case study may differ from the real CPS (e.g., autonomous vehicle), our simplified CPS represents a CPS in practice in terms of continuous interaction with the environment and distinction between different controller versions by multiple parameter values. Applying \textit{ENVI} to more complex CPS could show different results, but the applicability of \textit{ENVI} for various use cases (i.e., TOVK, TMVK, and TMVU) shown in this paper is still valid for CPSs with such characteristics. However, additional case studies with more complex CPS are required to improve our results' generalizability.

In terms of internal validity, the goal achievement measure defined based on specific driving quality metrics could be a potential threat since the evaluation of the lane-keeping system's goal could be biased to a specific aspect of driving. 
To mitigate this threat, in our case study, we defined eight driving qualities from the FOT logs motivated by \citet{cherrett2001extracting} and aggregated the results on the qualities to comprehensively understand whether the lane-keeping system under analysis works well or not. 
Hyperparameter value settings for machine learning models (e.g., number of iterations, learning rates, Etc.) could be another potential threat to the internal validity since the performance of machine learning models can largely depend on hyperparameter values~\cite{wang2018stealing, probst2019tunability}. 
We used the default values provided in the original studies~\cite{schulman2017proximal, ho2016generative}. 
Nevertheless, hyperparameter tuning is an important research field, so it remains an interesting future work.
In addition, the verification accuracy evaluation results could be affected by the simulation duration $T$ because small errors of the environment model can be accumulated and cause significant errors in a long simulation, as mentioned in Section \ref{sec:problemDef}. However, in our case study, we could not see the problem in all \textit{ENVI} algorithms even when $T$ is ten times longer than the current setting in this paper. Nevertheless, analyzing the performances on mitigating the compounding error of various IL algorithms for \textit{ENVI} in different systems remains an interesting future work.

\section{Discussion}\label{sec:discussion}

\textbf{\textit{IL algorithm selection}}: 
In this paper, we considered three representative IL algorithms (BC, GAIL, and BCxGAIL) for the environment model generation. In practice, a specific IL algorithm should be selected when implementing \textit{ENVI} considering its characteristics, as described in Section \ref{sec:background}. Based on our case study results, we recommend engineers use the BCxGAIL algorithm in practice since the environment models generated by BCxGAIL were the most accurate in terms of CPS goal verification. However, there are other factors for the IL algorithm selection, such as learning speed or sensitivity to hyperparameters, and therefore providing more empirical guidelines for selecting a specific IL algorithm still remains an interesting future work.

\textbf{\textit{Knowledge-based approach vs. Data-driven approach}}:
When there is a high-fidelity simulation engine based on well-known principles in the CPS domain, engineers can manually create an accurate virtual environment in the simulator for CPS goal verification. In contrast to such knowledge-based environment modeling, \textit{ENVI} is a data-driven approach where only a few FOT logs are required to automatically generate an accurate virtual environment model. This is a huge advantage when there are no high-fidelity simulators or well-defined principles in the CPS domain. Therefore, the data-driven approach can complement the knowledge-based approach depending on the application domain.

\textbf{\textit{Open challenges}}: Though we successfully developed and evaluated \textit{ENVI}, there are three main open challenges for data-driven environment model generation approaches.

First, sample efficiency is essential. This is because conducting FOTs to collect logs is the most expensive task in the data-driven approach. In our case study, BCxGAIL that combines BC and GAIL to improve sample efficiency indeed outperforms the other IL algorithms in most cases. Using state-of-the-art techniques for increasing sample efficiency~\cite{jena2020augmenting, zhang2020f, robertson2020concurrent} could further help.

Second, it should be robust to noise in FOT logs. We utilized IL techniques, and many IL studies assume the correctness of the expert demonstration~\cite{codevilla2018end, abdou2019end, peng2018deepmimic}. However, the demonstrator for IL algorithms in the data-driven environment model generation is the real environment, and therefore some level of noise can appear (e.g., due to sensor noise. In our case study, though we used noisy data collected by the real CPS, systematically investigating the impact of noise was not in the scope of our work. Nevertheless, as many studies have already considered the noise issue in machine learning~\cite{atla2011sensitivity,gupta2019dealing,zeng2021noise}, they could better guide how to address noisy data.

Third, finding a proper level of abstraction for complex environmental behaviors is important. We abstracted the environment as a state-transition function in a closed-loop simulation and recast the environment model generation problem as the IL problem (see Section \ref{sec:problemDef}). This is a typical level of abstraction when an environment is modeled~\cite{qin2016sit,reichstaller2018risk,shin2021concepts}. However, this simple representation may not be sufficient to model complex environmental behaviors, such as structural changes in the environment during the FOT or responses to factors other than the system. Therefore, an extension of the \textit{CPS-ENV interaction model} could be needed for some domains. It is an interesting future work, and we can also refer to some IL studies that imitate complex expert behaviors (e.g., multi-task or concurrent behavior)~\cite{agrawal2016task, harmer2018imitation, singh2020scalable}.

\section{Related work}\label{sec:relatedWork}

Instead of conducting FOTs, assessing CPS on a simulation environment is widely used in CPS engineering. Therefore, many studies have been presented in modeling CPS environments. 

\citet{qin2016sit} and \citet{reichstaller2018risk} modeled the interaction between the CPS and environment as a closed-loop similar to our CPS-ENV model. They generated environmental testing inputs to predict and evaluate the CPS runtime behavior. \citet{fredericks2016automatically} also specified uncertain situations the CPS may face at runtime, such as inaccurate or delayed cognition of the environment, to evaluate the CPS with adverse environmental inputs. These approaches can generate the initial environmental inputs that the CPS observes using sensors, but the state transition of the environment during the simulation should be manually modeled by domain experts or engineers in an external simulator.

Some studies model the environment state transition for CPS simulation similar to our approach. \citet{puschel2014combined} modeled the change of the environment state as a process model and reconfigured the environment based on the model during CPS simulation. \citet{yang2014verifying} explicitly specified how the environmental state is changed after CPS action on a state machine. \citet{camara2017reasoning} and \citet{moreno2018flexible} also modeled the probabilistic environment state transition in Markov Decision Process (MDP) and verified the CPS goal achievement in the dynamic environment. Though they modeled environmental state transition functions, domain experts have to manually design these models, which require sufficient domain knowledge and efforts. 

There are studies utilizing environmental data to model the environment. \citet{ding2015modeling} modeled the continuous environment state transition as a continuous place in an extension of Petri nets, and the parameters in the model were learned from data. \citet{aizawa2018identifying} and \citet{sykes2013learning} modeled the changing environment as a labeled transition system (LTS) and a logic program, respectively. The initial environment models are revised by execution trace data of the system so that the models represent the changing environment of reality as accurately as possible. However, in these studies, the revised environment model is still highly dependent on the initial models made by experts because data update only the partial information in the model. 

Unlike the previous studies that modeled the environment of CPS, we abstract the complex state transition of the environment into a black box function implemented as a neural network. As a result, the environment model can be automatically generated with execution trace samples of CPS without prior knowledge of the environment.

Independently from CPS, model-based Reinforcement Learning (RL) uses a notion of the environment model generally defined as anything that informs how the RL agent's environment will respond to the agent's actions~\cite{sutton2018reinforcement}. 
Though the concept is similar to our environment model that interacts with the CPS under verification, the purposes of training (learning) the environment model are different.
The primary objective of the environment model in model-based RL is to better learn the agent's policy function, so an inaccurate environment model is acceptable as long as it can promote the policy learning process. Naturally, supervised learning is used for learning the environment model~\cite{moerland2020model} without considering possible accumulations of errors over time. 
In contrast, the environment model in \textit{ENVI} is to replace the real FOT environment for CPS goal verification, and therefore making the environment model the same as the real environment in a closed-loop simulation is our primary objective, which is why we leverage Imitation Learning (IL) in our approach.

\section{Conclusion}\label{sec:conclusion}
In this paper, we present \textit{ENVI}, a novel data-driven environment imitation approach that efficiently generates accurate virtual environment models for CPS goal verification. Instead of conducting expensive FOTs many times, \textit{ENVI} requires only a few FOTs to collect some FOT logs for training a virtual environment model. By leveraging the representative IL algorithms (i.e., BC, GAIL, and BCxGAIL), an accurate virtual environment model can be generated automatically from the collected FOT logs. Our case study using a LEGO-lized autonomous vehicle equipped with a lane-keeping system shows that the CPS goal verification accuracy of the virtual environment models generated by our approach is very accurate, even when only a few FOT logs are used for training the models. The case study also shows that when the same CPS has multiple versions from an evolutionary development process, an \textit{ENVI}-generated environment model can be used for the CPS goal verification of new versions whose FOT logs are never collected before for the model training.

In future work, we plan to provide practical guidelines for using \textit{ENVI} with different IL algorithms by further investigating the characteristics of individual IL algorithms and conducting more case studies with complex CPS (e.g., an automated driving system composed of machine learning components). We further expect that \textit{ENVI} is not limited to the purpose of CPS controller verification, so we also plan to suggest new applications of \textit{ENVI}, such as an optimal CPS control predicting the environmental reaction.

\section*{Acknowledgements}
This research was supported by the MSIT (Ministry of Science and ICT), Korea, under the ITRC (Information Technology Research Center) support program (IITP-2022-2020-0-01795) and (SW Star Lab) Software R\&D for Model-based Analysis and Verification of Higher-order Large Complex System (No. 2015-0-00250) supervised by the IITP (Institute of Information \& Communications Technology Planning \& Evaluation).
This research was also partially supported by the Basic Science Research Program through the National Research Foundation of Korea (NRF) funded by the Ministry of Education (2019R1A6A-3A03033444).

\bibliography{bibliography}

\end{document}